\documentclass[12pt]{article}
\usepackage{a4wide}
\usepackage{graphicx}

\newcommand{\be}{\begin{equation}}
\newcommand{\ee}{\end{equation}}
\newcommand{\ba}{\begin{eqnarray}}
\newcommand{\ea}{\end{eqnarray}}
\newcommand{\mrm}[1]{\mathrm{#1}}

\newcommand{\nn}{\nonumber \\}
\newcommand{\Q}{{\cal Q}}

\begin{document}

\begin{titlepage}
\begin{flushright}
LU TP 04-20\\
hep-ph/0405025\\
May 2004\\
\end{flushright}
\vfill
\begin{center}
{\Large\bf Isospin Breaking in $K\to3\pi$ Decays I: \\ 
\vspace{0.5 cm}
Strong Isospin Breaking}

\vfill

{\bf Johan Bijnens and Fredrik Borg}\\[1cm]
{Department of Theoretical Physics, Lund University\\
S\"olvegatan 14A, S 22362 Lund, Sweden}
\end{center}

\vfill

\begin{abstract}
The CP conserving amplitudes for the
decays $K\to3\pi$ are calculated in Chiral Perturbation 
Theory. The calculation is made at the next-to-leading order including 
strong and local electromagnetic
isospin breaking. A comparison is made between the squared amplitudes
with and without isospin violation to estimate the size of the effect. 
We find corrections of order five percent in the amplitudes.  
\end{abstract}
\vfill

\end{titlepage}

\section{Introduction}
\label{introduction}

Chiral Perturbation Theory (ChPT) is the effective field theory
of the low-energy strong interactions. It was introduced in its present form
by Weinberg, Gasser and Leutwyler \cite{Weinberg,GL1,GL2} and it has
had many successes and applications. A pedagogical introduction
can be found in \cite{chptlectures}.
The theory has been applied as well for nonleptonic weak decays. The
main work of extending ChPT to the nonleptonic weak interaction was 
done by Kambor, Missimer and Wyler 
who worked out the general formalism \cite{KMW1} and applied it
to $K\to2\pi,3\pi$ decays \cite{KMW2}. These results were then used
to obtain directly relations between physical observables in
\cite{KDHMW}. Reviews of applications of ChPT to nonleptonic
weak interactions are \cite{chptweakreviews}. The expressions of
Ref.~\cite{KMW2} were never published and have been lost.

The next-to-leading order 
$K\to2\pi$ amplitudes in the isospin limit were recalculated in \cite{BPP,BDP}
and the $K\to3\pi$ amplitudes in \cite{BDP}. The latter have been
confirmed in \cite{GPS}.
In \cite{BDP} a full fit to all available $K\to2\pi$ and $K\to3\pi$
experimental results was performed and it was found that one could fit 
the decay rates and the slopes in the Dalitz plot. There was a discrepancy
with the observed quadratic slopes or curvatures in the Dalitz plot and
this can have several different origins. It could be an experimental
problem (especially given the discrepancies between several experiments) or it
could have a theoretical origin. In the latter the amplitudes calculated
so far have three different types of corrections, isospin breaking,
electromagnetic corrections or higher order ChPT corrections.
In this paper we investigate the first class. Work is in progress to evaluate
the electromagnetic corrections as well. More precisely, in this paper we
investigate effects of the quark mass difference $m_u-m_d$ as well as the
local electromagnetic effects. We have worked out all orders in $m_u-m_d$
to order four in the chiral expansion but we have found that the numerical
difference between the all order isospin breaking and the first order
effect was much smaller than the other uncertainties in the calculations.
We therefore present results only to first order in $m_u-m_d$.

There exists very little work on isospin breaking in $K\to3\pi$ decays
and none that we know of within ChPT. The $K\to2\pi$ case has been
investigated more extensively since it has possibly strong
effects on $\epsilon^\prime/\epsilon$~\cite{penguinomega}. Recent works
are \cite{GV,ENP,WM,CDG}.

The outline of this paper is as follows. The next section describes strong 
isospin breaking. Section \ref{lagrangian} presents the Chiral 
Lagrangians needed for the calculation
to next-to-leading order and section \ref{kinematics} specifies the decays and 
describes the relevant kinematics. The analytical results, the amplitudes for 
$K\to3\pi$, are described in section \ref{isoanalytres} and in section 
\ref{isonumres} some numerical results are shown. 
The last section contains the conclusions.  

\section{Strong Isospin Breaking}
\label{striso}

Historically, the isospin approximation meant treating the proton and neutron 
in the same way. In a quark picture this corresponds to treating the up-
and down-quark as being identical. This implies putting $m_u=m_d$ as well as
neglecting electromagnetism. Obviously this is an approximation but since
one can roughly estimate the corrections, and it greatly 
simplifies the calculations, its use is very common. 
However, to reduce the
errors one has to quote, the isospin approximation has to be abandoned. 

Isospin breaking is usually divided into two parts. Strong isospin breaking
coming from the fact that $m_u\neq m_d$, and electromagnetic isospin breaking
from including electromagnetic effects. In this paper we only take into account
the strong part and the local part (the part which doesn't include explicit
photons) of the EM isospin breaking. From now on we will use the name 
strong isospin breaking for the sum of these two effects.

The difference between $m_u$ and $m_d$ leads to mixing between $\pi$ and
$\eta$. This means changes in the formulas for both the physical masses of
$\pi$ and $\eta$ as well as the amplitude for any process involving either 
of the two. For a detailed discussion see \cite{strongiso}.

Including the local EM part means introducing new Lagrangians at each order,
 proportional to $e^2$ and $p^2\,e^2$ respectively.    

\section{The ChPT Lagrangians}
\label{lagrangian}

The starting point of our ChPT calculation is the Lagrangians. The 
order parameters in the perturbation series are $p$, the momenta of, 
and $m$, the 
mass of the pseudoscalars. Including isospin breaking also $e$, the electron
charge, is used as an order parameter. Leading order (order two) therefore 
means terms of order $p^2,m^2,e^2$, and next-to-leading order 
(order four) $p^4,p^2\,m^2,m^4,p^2\,e^2$ and $m^2\,e^2$ ($e^4$ is neglected). 
 
\subsection{Leading Order}
\label{leading}

Our leading order effective Lagrangian assumes the form
\be
{\cal L}_{2}={\cal L}_{S2}+{\cal L}_{W2}+{\cal L}_{E2}
\ee
Here ${\cal L}_{S2}$ refers to the strong $\Delta S = 0$ part, ${\cal L}_{W2}$ 
the weak 
$\Delta S = \pm 1$ part, and ${\cal L}_{E2}$ the strong-electromagnetic and 
weak-electromagnetic parts combined. For the strong part we have \cite{GL1}
\be
\label{L2S}
{\cal L}_{S2} = \frac{F_0^2}{4} \; \langle u_\mu u^\mu + \chi_+\rangle
\ee
Here $\langle A\rangle$ stands for the flavour trace of the matrix $A$,
and $F_0$ is the pion decay constant in the chiral limit.
We also define the matrices $u_\mu$, $u$ and $\chi_\pm$
\be
u_\mu = i u^\dagger\, D_\mu U\, u^\dagger = u_\mu^\dagger\,,\quad u^2 = U\,,
\quad \chi_\pm = u^\dagger \chi u^\dagger \pm u \chi^\dagger u\,,
\ee
where the special unitary matrix $U$ contains the Goldstone boson fields
\be
U = \exp\left(\frac{i\sqrt{2}}{F_0}M\right)\,,\quad
M =\left(\begin{array}{ccc}
\frac{1}{\sqrt{2}}\pi_3+\frac{1}{\sqrt{6}}\eta_8 & \pi^+ & K^+\\
\pi^- & \frac{-1}{\sqrt{2}}\pi_3+\frac{1}{\sqrt{6}}\eta_8 & K^0\\
K^- & \overline{K^0} & \frac{-2}{\sqrt{6}}\eta_8
         \end{array}\right)\,.
\ee
The formalism we use is the external field method of \cite{GL1}, 
but for our purpose it suffices to set
\be
\chi = 2 B_0
\left(\begin{array}{ccc}m_u &  & \\ & m_d & \\ & & m_s\end{array}\right)\, 
\,\,\,\, \mrm{and}\,\,\,\, D_\mu U = \partial_\mu U.
\ee
We diagonalize the quadratic terms in (\ref{L2S}) by a rotation
\ba
\pi^0 &=& \pi_3\cos\epsilon + \eta_8\sin\epsilon\,
\nonumber\\
\eta &=& -\pi_3\sin\epsilon + \eta_8\cos\epsilon\,,
\ea
where the lowest order mixing angle $\epsilon$ satisfies
\be
\tan(2\epsilon) = \frac{\sqrt{3}}{2}\frac{m_d-m_u}{m_s-\hat m}\,,
\ee
with $2 \hat m = m_u+m_d$.

For the weak part the Lagrangian has the form \cite{Cronin}
\be
\label{LW2}
{\cal L}_{W2} = C \, F_0^4 \, 
\Bigg[ G_8 \langle \Delta_{32} u_\mu u^\mu \rangle +
G_8' \langle\Delta_{32} \chi_+ \rangle  +
G_{27} t^{ij,kl} \, \langle \Delta_{ij } u_\mu \rangle
\langle\Delta_{kl} u^\mu \rangle \Bigg]
+ \mbox{ h.c.}\,.\nonumber
\ee
The tensor $t^{ij,kl}$ has as nonzero components
\ba
\label{deft}
t^{21,13} =
t^{13,21} = \frac{1}{3} \, &;& \, 
t^{22,23}=t^{23,22}=-\frac{1}{6} \, ; \nonumber \\
t^{23,33}=t^{33,23}=-\frac{1}{6} \, &;& \, 
t^{23,11} =t^{11,23}=\frac{1}{3}\,,
\ea
and the matrix $\Delta_{ij}$ is defined as
\be
\Delta_{ij} \equiv u \lambda_{ij} u^\dagger\,,\quad
\left(\lambda_{ij}\right)_{ab} \equiv \delta_{ia} \, \delta_{jb}\,.
\ee
The coefficient $C$ is defined such that in the chiral
and large $N_c$ limits $G_8 = G_{27} =1$,
\be
C= -\frac{3}{5} \, \frac{G_F}{\sqrt 2} V_{ud} \, V_{us}^* \, .
\ee

Finally, the remaining electromagnetic part, relevant for this calculation,
 looks like (see e.g. \cite{EMEcker})
\be
\label{LE2}
{\cal L}_{E2} = 
e^2 F_0^4 Z \langle {\cal Q}_L {\cal Q}_R\rangle +
e^2 F_0^4 \langle \Upsilon {\cal Q}_R\rangle
\ee
where the strangeness-conserving term includes
\begin{equation} \label{defZ}
\Q_L = uQu^\dagger\,,\quad \Q_R = u^\dagger Qu\,\,\,\, \mrm{with}\,\,\,\,
Q = \left[ \begin{array}{ccc} 2/3 & 0 & 0 \\ 0 & -1/3 & 0 \\ 0 & 0 & -1/3
\end{array}
\right],
\ee
and the weak-electromagnetic term is characterized by a 
constant $G_E$ \mbox{($G_E=g_{\rm ewk}G_8$ in \cite{EMEcker})},
\begin{equation} \label{gewk}
\Upsilon = G_E\, F_0^2 \Delta_{32} + {\rm h.c.}  \, .
\ee

\subsection{Next-to-leading Order}
\label{ntleading}

Since ChPT is a non-renormalizable theory, new terms have to be added at 
each order to compensate for the loop-divergences. This means that
the Lagrangians increase in size for every new order and the number of 
free parameters rise as well.
At next-to-leading order we split the Lagrangian in four parts which, in
obvious notation, are 
\be
{\cal L}_{4}={\cal L}_{S4}+{\cal L}_{W4}+{\cal L}_{S2E2}+{\cal L}_{W2E2}(G_8)
\,.
\ee
Here the notation $(G_8)$ indicates that here only the dominant
$G_8$-part is included in the Lagrangian, and therefore in the calculation.
The strong part we need looks like \cite{GL1, radiative}
 \ba
\label{LS4}
{\cal L}_{S4}&=&
L_1 \; \langle u_\mu u^\mu \rangle^2 + L_2 \; \langle u_\mu
u^\nu\rangle \; \langle u^\mu u_\nu\rangle 
+ L_3 \; \langle u_\mu u^\mu u_\nu u^\nu\rangle + 
L_4 \; \langle u_\mu u^\mu\rangle \; \langle \chi_+\rangle \nn 
&& \mbox{} + L_5 \; \langle u_\mu u^\mu \chi_+\rangle + 
L_6 \; \langle \chi_+\rangle^2 + L_7 \; \langle \chi_-\rangle^2 
+  L_8 \; \frac{1}{2} \left(\langle \chi_+^2\rangle +
\langle \chi_-^2\rangle\,\right),
\ea
and the weak part, quoting only the terms
relevant for $K\to3\pi$ decays, is 
\cite{KMW1,Esposito,EKW},
\ba
\label{LW4}
{\cal L}_{W4} &=& C \, F_0^2 \, \Bigg\{G_8\Big[
N_1 {\cal O}^8_1 + N_2 {\cal O}^8_2 + N_3 {\cal O}^8_3 + 
N_4 {\cal O}^8_4 +
N_5 {\cal O}^8_5 + N_{6} {\cal O}^8_{6} 
+ N_{7} {\cal O}^8_{7}
 \nonumber \\&&
 + N_{8} {\cal O}^8_{8} + N_{9} {\cal O}^8_{9} 
+ N_{10} {\cal O}^8_{10} + N_{11} {\cal O}^8_{11} + 
N_{12} {\cal O}^8_{12} + N_{13} {\cal O}^8_{13} \Big]
\nonumber \\&&
+  G_{27} \,\Big[
D_1 {\cal O}^{27}_1 + D_2 {\cal O}^{27}_2 +
D_{4} {\cal O}^{27}_4 + D_{5} {\cal O}^{27}_5 
 + D_{6} {\cal O}^{27}_6 + D_{7} {\cal O}^{27}_7 + D_{26} {\cal O}^{27}_{26}
\nonumber \\&& 
+ D_{27} {\cal O}^{27}_{27} + D_{28} {\cal O}^{27}_{28}
+  D_{29} {\cal O}^{27}_{29}
+ D_{30} {\cal O}^{27}_{30} + D_{31} {\cal O}^{27}_{31} \Big]\Bigg\}
+ \mbox{h.c.}\,. 
\ea
The octet operators are

\be
\begin{array}{lccl}
{\cal O}^8_1 = \langle\Delta_{32} u_{\mu} u^{\mu} u_{\nu} u^{\nu} \rangle
&&&
{\cal O}^8_2 = \langle\Delta_{32} u_{\mu} u_{\nu} u^{\nu} u^{\mu} \rangle
\nn [0.2cm]
{\cal O}^8_3 = \langle\Delta_{32} u_{\mu} u_{\nu}\rangle\langle u^{\mu} 
u^{\nu} \rangle
&&&
{\cal O}^8_4 = \langle\Delta_{32} u_{\mu}\rangle\langle u_{\nu} u^{\mu} 
u^{\nu} \rangle
\nn [0.2cm]
{\cal O}^8_5 
= \langle\Delta_{32} (\chi_+ u_{\mu} u^{\mu}+u_{\mu} u^{\mu} \chi_+) 
\rangle
&&&
{\cal O}^8_6 
= \langle\Delta_{32} u_{\mu}\rangle\langle u^{\mu} \chi_+ \rangle 
\nn [0.2cm]
{\cal O}^8_7 
= \langle\Delta_{32} \chi_+\rangle\langle u_{\mu} u^{\mu} \rangle 
&&&
{\cal O}^8_8 
= \langle\Delta_{32} u_{\mu} u^{\mu}\rangle\langle\chi_+ \rangle 
\nn [0.2cm]
{\cal O}^8_9 
= \langle\Delta_{32} (\chi_- u_{\mu} u^{\mu}-u_{\mu} u^{\mu} \chi_-)
 \rangle 
&&&
{\cal O}^8_{10} 
= \langle\Delta_{32} \chi_+ \chi_+ \rangle 
\nn [0.2cm]
{\cal O}^8_{11} 
= \langle\Delta_{32} \chi_+\rangle\langle\chi_+ \rangle
&&&
{\cal O}^8_{12} 
= \langle\Delta_{32} \chi_- \chi_- \rangle 
\nn [0.2cm]
{\cal O}^8_{13} 
= \langle\Delta_{32} \chi_-\rangle\langle\chi_- \rangle \,,
&&&
\end{array}
\ee
\vspace{0cm}\\
and the 27 operators are
\ba
{\cal O}^{27}_1 
&=& t^{ij,kl}  \langle \Delta_{ij} \chi_+\rangle
\langle\Delta_{kl} \chi_+\rangle \,
\nonumber \\
{\cal O}^{27}_2 
&=& t^{ij,kl}  \langle\Delta_{ij} \chi_-\rangle
\langle\Delta_{kl} \chi_- \rangle \,
\nonumber \\
{\cal O}^{27}_4 
&=& t^{ij,kl}  \langle\Delta_{ij} u_{\mu}\rangle
\langle\Delta_{kl} (u^{\mu} \chi_++\chi_+ u^{\mu}) \rangle \,
\nonumber \\
{\cal O}^{27}_5 
&=& t^{ij,kl}  \langle\Delta_{ij} u_{\mu}\rangle
\langle\Delta_{kl} (u^{\mu} \chi_--\chi_- u^{\mu}) \rangle\, 
\nonumber \\
{\cal O}^{27}_6 
&=& t^{ij,kl}  \langle\Delta_{ij} \chi_+\rangle
\langle\Delta_{kl} u_{\mu} u^{\mu} \rangle \,
\nonumber \\
{\cal O}^{27}_7 
&=& t^{ij,kl}  \langle\Delta_{ij} u_{\mu}\rangle
\langle\Delta_{kl} u^{\mu}\rangle\langle\chi_+ \rangle \,
\nonumber \\
{\cal O}^{27}_{26} 
&=& t^{ij,kl}  \langle\Delta_{ij} u_{\mu} u^{\mu}\rangle
\langle\Delta_{kl} u_{\nu} u^{\nu} \rangle\,
\nonumber \\
{\cal O}^{27}_{27} 
&=& t^{ij,kl}  \langle\Delta_{ij} (u_{\mu} u_{\nu}+u_{\nu} u_{\mu})\rangle
\langle\Delta_{kl} u^{\mu} u^{\nu} \rangle\,
\nonumber \\
{\cal O}^{27}_{28} 
&=& t^{ij,kl}  \langle\Delta_{ij} (u_{\mu} u_{\nu}-u_{\nu} u_{\mu})\rangle
\langle \Delta_{kl} u^{\mu} u^{\nu} \rangle\,
\nonumber \\
{\cal O}^{27}_{29} 
&=& t^{ij,kl}  \langle\Delta_{ij} u_{\mu}\rangle
\langle\Delta_{kl} u_{\nu} u^{\mu} u^{\nu} \rangle\,
\nonumber \\
{\cal O}^{27}_{30} 
&=& t^{ij,kl}  \langle\Delta_{ij} u_{\mu}\rangle
\langle\Delta_{kl} (u^{\mu} u_{\nu} u^{\nu}+u_{\nu} u^{\nu} u^{\mu}) \rangle\,
\nonumber \\
{\cal O}^{27}_{31} 
&=& t^{ij,kl}  \langle\Delta_{ij} u_{\mu}\rangle
\langle\Delta_{kl} u^{\mu}\rangle\langle u_{\nu} u^{\nu} \rangle\,.
\ea

The complete minimal Lagrangian of ${\cal O}(G_8 e^2 p^2)$
takes the form \cite{EMEcker}
\be 
\label{W2E2}
{\cal L}_{W2E2}(G_8) =  G_8 e^2 F_0^4 \sum_{i=1}^{14} Z_i Q^{w}_i 
+ \mbox{ h.c.}
\ee 
with operators $Q_i^w$ of ${\cal O}(p^2)$ and dimensionless coupling 
constants $Z_i$.
A linear independent set of  operators  is given by
\be
\begin{array}{lccl}
Q^{w}_1 = \langle \Delta_{32} \{ \Q_R, \chi_+ \} \rangle
&&&
Q^{w}_2 = \langle \Delta_{32} \Q_R \rangle \langle \chi_+ \rangle 
\nn [0.2cm]
Q^{w}_3 = \langle \Delta_{32} \Q_R \rangle \langle \chi_+ \Q_R \rangle 
&&&
Q^{w}_4 = \langle \Delta_{32} \chi_+ \rangle \langle \Q_L \Q_R \rangle 
\nn [0.2cm]
Q^{w}_5 = \langle \Delta_{32} u_\mu u^\mu \rangle
&&&
Q^{w}_6 = \langle \Delta_{32} \{\Q_R,u_\mu u^\mu\} \rangle
\nn [0.2cm]
Q^{w}_7 = \langle \Delta_{32} u_\mu u^\mu \rangle \langle \Q_L \Q_R \rangle 
&&&
Q^{w}_8 = \langle \Delta_{32} u_\mu \rangle \langle \Q_L u^\mu \rangle
\nn [0.2cm]
Q^{w}_9 = \langle \Delta_{32} u_\mu \rangle \langle \Q_R u^\mu \rangle
&&&
Q^{w}_{10} = \langle \Delta_{32} u_\mu \rangle \langle \{\Q_L,\Q_R\}u^\mu
\rangle
\nn [0.2cm]
Q^{w}_{11} = \langle \Delta_{32} \{\Q_R,u_\mu\} \rangle \langle \Q_L u^\mu 
\rangle
&&&
Q^{w}_{12} = \langle \Delta_{32} \{\Q_R,u_\mu\} \rangle \langle \Q_R u^\mu 
\rangle
\nn [0.2cm]
Q^{w}_{13} = \langle \Delta_{32} \Q_R \rangle \langle u_\mu u^\mu \rangle
&&&
Q^{w}_{14} = \langle \Delta_{32} \Q_R \rangle \langle u_\mu u^\mu \Q_R 
\rangle~.
\end{array}
\label{list1}
\ee
\vspace{0cm}\\
Finally the strong-electromagnetic part, which looks like 
\cite{radiative,urech}
\be 
\label{S2E2}
{\cal L}_{S2E2} =  e^2 F_0^2 \sum_{i=1}^{11} K_i Q^{s}_i 
\ee 
with
\be
\begin{array}{lccl}
Q^{s}_1 = \frac{1}{2} \; \langle \Q_{\rm L}^2 +
\Q_{\rm R}^2\rangle \; \langle u_\mu u^\mu\rangle   
& & &
Q^{s}_2 = \langle \Q_{\rm L} \Q_{\rm R}\rangle 
\; \langle u_\mu u^\mu \rangle \nn [0.2cm]
Q^{s}_3 = (\langle \Q_{\rm L} u_\mu\rangle 
\; \langle \Q_{\rm L} u^\mu
\rangle + \langle \Q_{\rm R} u_\mu\rangle 
\; \langle \Q_{\rm R} u^\mu\rangle )
& & &
Q^{s}_4 = \langle \Q_{\rm L} u_\mu\rangle 
\; \langle \Q_{\rm R} u^\mu \rangle \nn [0.2cm]
Q^{s}_5 = \langle(\Q_{\rm L}^2 + \Q_{\rm R}^2)\, u_\mu u^\mu\rangle 
&&&
Q^{s}_6 = \langle (\Q_{\rm L} \Q_{\rm R} + 
\Q_{\rm R} \Q_{\rm L})\, u_\mu u^\mu\rangle \nn [0.2cm]
Q^{s}_7 = \frac{1}{2} \; \langle \Q_{\rm L}^2 
+ \Q_{\rm R}^2\rangle \; \langle \chi_+\rangle 
&&&
Q^{s}_8 = \langle \Q_{\rm L} \Q_{\rm R}\rangle 
\; \langle \chi_+\rangle \nn [0.2cm]
Q^{s}_9 = \langle (\Q_{\rm L}^2 + \Q_{\rm R}^2) \chi_+\rangle 
&&&
Q^{s}_{10} = \langle(\Q_{\rm L} \Q_{\rm R} 
+ \Q_{\rm R} \Q_{\rm L}) \chi_+\rangle \nn [0.2cm]
Q^{s}_{11} = \langle(\Q_{\rm R} \Q_{\rm L} 
- \Q_{\rm L} \Q_{\rm R}) \chi_-\rangle\,. 
&&&
\end{array}
\label{list2}
\ee

The infinities appearing in the loop diagrams are canceled by
replacing the coefficients in (\ref{LS4}), (\ref{LW4}), (\ref{W2E2}) and 
(\ref{S2E2}) by
the renormalized coefficients and a subtraction part.
The infinities needed in the strong sector were calculated first
in \cite{GL2} and those for the weak sector
in \cite{KMW1}.
The terms are all of the type
\be
\label{definf}
X_i = (e^c\mu)^{-2\epsilon}
\left(X_i^r + x_i\frac{1}{16\pi^2 \epsilon}\right)\,,
\ee
with the dimension of space-time $d=4-2\epsilon$ and
\be
c = -\frac{1}{2}\left(\ln(4\pi) +\Gamma^\prime(1)+1\right)\,.
\ee
For $X=L,N,D $ see \cite{BDP}, and for $X=Z,K $ see Table \ref{tabinf}
\cite{EMEcker,urech}. However, since the full $e^2p^2$-contribution 
 is not included (the explicit photon diagrams are left out), 
the infinities don't cancel completely. They still provide a 
number of checks on the calculations though. Since contributions of order 
$e^4$ are neglected, Lagrangians allready containing $e^2$ will not 
add any explicit photon terms.  Therefore the infinities 
proportional to $Z$ and $G_E$ 
cancel. Even more, since there are no explicit photon contributions to the 
decays involving only neutral particles, like $K_L~\to~\pi^0  \pi^0 \pi^0$, 
the infinities in that channel cancel completely.
\begin{table}
\begin{center}
\begin{tabular}{|c|c|c|c|}
\hline
$Z_i$   &   $z_i$  &   $K_i$   &  $k_i$   \\
\hline
1       & $17/24 +3/2\,Z -3/4\,g_{\rm ewk} $ 
        & 1  &  $-3/8  $                           \\
2       & $-1/2 -8/3\,Z -1/2\,g_{\rm ewk} $   
        & 2  &  $-1/2\,Z $                             \\
3       & $-3/8 -7/2\,Z      $   
        & 3  &  $3/8   $                               \\
4       & $3/8 +7/2\,Z      $   
        & 4  &  $-Z$                           \\
5       & $1$   
        & 5  &  $9/8$  \\
6       & $-7/4 -5/2\,Z -3/4\,g_{\rm ewk}$   
        & 6  &  $-3/4\,Z$                              \\
7       & $-3/4 -5/2\,Z      $   
        & 7  &  $0$  \\
8       & $1/4  $   
        & 8  &  $-1/2\,Z  $                            \\
9       & $11/12 -2/3\,Z - g_{\rm ewk}   $   
        & 9  &  $1/8$  \\
10      & $3/4 +1/2\,Z    $   
        &       10 &  $-1/8 -3/4\,Z$  \\
11      & $3/4 +Z    $                
        & 11 &  $-1/16$  \\
12      & $-3/4$              
        &    &    \\
13      & $35/24 +3/2\,Z -1/2\,g_{\rm ewk} $            
        &    &                           \\
14      & $-3/2 -15/2\,Z    $               &    &    \\
\hline
\end{tabular}
\end{center}
\caption{The coefficients of the subtraction of the infinite
parts defined in Eq. (\ref{definf}).\label{tabinf}}
\end{table}

\section{Kinematics}
\label{kinematics}

There are five different CP-conserving decays of the type $K\to3\pi$:
\ba
\label{defdecays}
K_L(k)&\to&\pi^0(p_1)\,\pi^0(p_2)\,\pi^0(p_3)\,,\quad [A^L_{000}]\,,\nonumber\\
K_L(k)&\to&\pi^+(p_1)\,\pi^-(p_2)\,\pi^0(p_3)\,,\quad [A^L_{+-0}]\,,\nonumber\\
K_S(k)&\to&\pi^+(p_1)\,\pi^-(p_2)\,\pi^0(p_3)\,,\quad [A^S_{+-0}]\,,\nonumber\\
K^+(k)&\to&\pi^0(p_1)\,\pi^0(p_2)\,\pi^+(p_3)\,,\quad [A_{00+}]\,,\nonumber\\
K^+(k)&\to&\pi^+(p_1)\,\pi^+(p_2)\,\pi^-(p_3)\,,\quad [A_{++-}]\,,
\ea
where we have indicated the four-momentum defined for each particle
and the symbol used for the amplitude.
The $K^-$ decays are not treated separately since they are counterparts
to the $K^+$ decays.

The kinematics is treated using
\be
s_1 = \left(k-p_1\right)^2\,,\quad
s_2 = \left(k-p_2\right)^2\,,\quad
s_3 = \left(k-p_3\right)^2\,.
\ee
The amplitudes are expanded in terms of the Dalitz plot variables
$x$ and $y$ defined as
\be
\label{defsi}
y =\frac{ s_3-s_0}{m_{\pi^+}^2}\,,\quad
x =\frac{ s_2-s_1}{m_{\pi^+}^2}\,,\quad
s_0 = \frac{1}{3}\left(s_1+s_2+s_3\right)\,.
\ee
The amplitudes for
 $K_L\to\pi^+\pi^-\pi^0,K^+\to\pi^+\pi^+\pi^-$ and $K^+\to\pi^0\pi^0\pi^+$
are symmetric under the interchange of the first two pions because of
CP or Bose-symmetry. The amplitude for  $K_L\to\pi^0\pi^0\pi^0$ is of course 
symmetric
under the interchange of all three final state particles and the one for
$K_S\to\pi^+\pi^-\pi^0$ is antisymmetric under the interchange of $\pi^+$ and
$\pi^-$ because of CP.

The amplitudes to order four can be written in terms of
single variable functions $M_i(s)$ \cite{BDP} as
\ba
\label{isodefMi}
A^L_{000}(s_1,s_2,s_3) &=& M_0(s_1)+M_0(s_2)+M_0(s_3) \,,
\nonumber\\
A^L_{+-0}(s_1,s_2,s_3) &=& M_1(s_3)+M_2(s_1)+M_2(s_2)+M_3(s_1)(s_2-s_3)
+M_3(s_2)(s_1-s_3)
\,,\nonumber\\
A^S_{+-0}(s_1,s_2,s_3) &=& M_4(s_1)-M_4(s_2)+M_5(s_1)(s_2-s_3)
-M_5(s_2)(s_1-s_3)
\nonumber\\&&
+M_6(s_3)(s_1-s_2)\,,
\nonumber\\
A_{00+}(s_1,s_2,s_3) &=&  M_7(s_3)+M_8(s_1)+M_8(s_2)+M_9(s_1)(s_2-s_3)
+M_9(s_2)(s_1-s_3)\,,
\nonumber\\
A_{++-}(s_1,s_2,s_3) &=&  M_{10}(s_3)+M_{11}(s_1)+M_{11}(s_2)
+M_{12}(s_1)(s_2-s_3)
+M_{12}(s_2)(s_1-s_3) \,.
\nonumber\\
\ea
The functions $M_i(s)$ are not unique. The full list of ambiguities
can be found in App.~A of \cite{BDP}.

\section{Analytical Results}
\label{isoanalytres}

\subsection{Lowest order}

The two diagrams that contribute to lowest order can be seen in
Fig.\ref{isofigtree}.
\begin{figure}
\begin{center}
\includegraphics{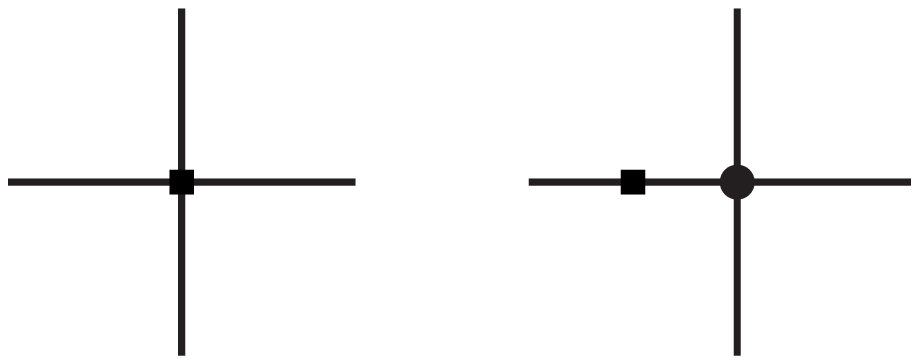}
\end{center}
\caption{The tree level diagrams for $K\to3\pi$. A filled square is a vertex
from ${\cal L}_{W2}$ or ${\cal L}_{E2} (\Delta S = 1)$ and a filled circle 
from ${\cal L}_{S2}$ or ${\cal L}_{E2} (\Delta S = 0)$.
\label{isofigtree}}
\end{figure}

One of the problems in expressing the amplitudes is the various possible
choices of masses to express the isospin conserving piece in.
The simplest expressions usually follow when using a mixture of
the neutral and charged masses and in the part that is obviously isospin
violating we can use the neutral and charged masses interchangeably, 
denoted by $m_{K}^2$ and $m_{\pi}^2$. 

The simplest expressions for the lowest order we have found are
\be
M_0(s)|_{p^2} = i\,\frac{C\, F_0^4}{F_\pi^3\, F_K}
    \left( G_8 - G_{27}\right)\,
    \frac{1}{3}\, m_{K^0}^2 \,\left(1+\sqrt{3}\,\sin\epsilon\right) \,.
\ee

\ba
M_1(s)|_{p^2} &=& i\frac{C F_0^4}{F_\pi^3 F_K}
   \Bigg\{\left( G_8 - G_{27}\right)
    \frac{1}{3} m_{K^0}^2
   +\frac{\sin\epsilon}{\sqrt3}\left(G_8-\frac{13}{3}G_{27}\right) m_K^2
\nonumber\\&&
   + Z e^2 F_0^2\left(\frac{4}{3} G_8 
    + \frac{G_{27}}{3}\frac{6 m_K^2-16 m_\pi^2}{m_K^2-m_\pi^2}\right)\Bigg\}\,,
\nonumber\\
M_2(s)|_{p^2} &=& 0\,,
\nonumber\\
M_3(s)|_{p^2} &=&i\frac{C F_0^4}{F_\pi^3 F_K}
   \Bigg\{-\frac{G_8}{3}+G_{27}\left(-\frac{1}{2}
    +\frac{1}{3}\frac{m_{\pi^+}^2}{m_{K^+}^2-m_{\pi^+}^2}
    +\frac{1}{2}\frac{m_{\pi^0}^2}{m_{K^0}^2-m_{\pi^0}^2}\right)
\nonumber\\&&
    + \frac{\sin\epsilon}{\sqrt{3}}\left(-2 G_8
        +\frac{17}{6}G_{27}\frac{m_K^2}{m_K^2-m_\pi^2}\right)
    + \left(Z\,G_8+\frac{1}{2} G_E\right)\frac{e^2 F_0^2}{m_K^2-m_\pi^2}
 \Bigg\}.
\ea

\ba
M_4(s)|_{p^2} &=& 0\,,
\nonumber\\
M_5(s)|_{p^2} &=& 0\,,
\nonumber\\
M_6(s)|_{p^2} &=&i\frac{C F_0^4}{F_\pi^3 F_K}
\Bigg\{
     G_{27}\left(-\frac{5}{2}
    -\frac{1}{3}\frac{m_{\pi^+}^2}{m_{K^+}^2-m_{\pi^+}^2}
    -\frac{1}{2}\frac{m_{\pi^0}^2}{m_{K^0}^2-m_{\pi^0}^2}\right)
\nonumber\\&&
   +\frac{\sin\epsilon}{\sqrt{3}}
    \left(-G_8+\frac{G_{27}}{2}\frac{7 m_K^2-6 m_\pi^2}{m_K^2-m_\pi^2}\right)
   +\left(-Z\,G_8 -\frac{1}{2} G_E\right)\frac{e^2 F_0^2}
{m_K^2-m_\pi^2}\Bigg\}\,.
\ea
This is in the isospin limit the same as (31) in \cite{BDP}
using Eq.~(A.9) there.

\ba
M_7(s)|_{p^2} &=&i\frac{C F_0^4}{F_\pi^3 F_K}
\Bigg\{\left(-\frac{1}{3}G_8-\frac{2}{9}G_{27}\right) m_{K^+}^2
   -\frac{10}{3}\frac{\sin\epsilon}{\sqrt{3}} m_K^2 G_{27}
\nonumber\\&&
   + Z e^2 F_0^2\left(-\frac{8}{3} G_8+\frac{G_{27}}{9}
              \frac{-46 m_K^2+16 m_\pi^2}{m_K^2-m_\pi^2}\right)
   - G_E e^2 F_0^2\Bigg\}\,,
\nonumber\\
M_8(s)|_{p^2} &=& 0\,,
\nonumber\\
M_9(s)|_{p^2} &=&i\frac{C F_0^4}{F_\pi^3 F_K}
\Bigg\{
    \frac{G_8}{3}+ G_{27}\left(\frac{19}{18}
    +\frac{1}{3}\frac{m_{\pi^+}^2}{m_{K^+}^2-m_{\pi^+}^2}
    +\frac{1}{2}\frac{m_{\pi^0}^2}{m_{K^0}^2-m_{\pi^0}^2}\right)
\nonumber\\&&
   +\frac{\sin\epsilon}{\sqrt{3}}
    \left(G_8+\frac{G_{27}}{3}\frac{7 m_K^2-6 m_\pi^2}{m_K^2-m_\pi^2}\right)
   +\left(Z\,G_8 +\frac{1}{2} G_E\right)\frac{e^2 F_0^2}
{m_K^2-m_\pi^2}\Bigg\}\,.
\ea
This can be brought into the form of (32) of \cite{BDP} using the
equivalents of (A.3-A.5) there.
\ba
M_{10}(s)|_{p^2} &=&i\frac{C F_0^4}{F_\pi^3 F_K}
\Bigg\{
G_8\left(s-m_{K^+}^2-m_{\pi^+}^2\right)
+G_{27}\left(-\frac{13}{3} s + m_{K^+}^2 +\frac{13}{3} m_{\pi^+}^2\right)
\nonumber\\&&
- 8 Z e^2 F_0^2\, G_8 -\frac{16}{3} Z e^2 F_0^2\, G_{27} -2 \,G_E e^2 F_0^2
\Bigg\}\,,
\nonumber\\
M_{11}(s)|_{p^2} &=& 0\,,
\nonumber\\
M_{12}(s)|_{p^2} &=& 0\,.
\ea
Here $F_\pi$ and $F_K$ are the pion and kaon decay constants 
respectively. 

The terms proportional to $G_E$ are published before in \cite{GPS} and,
after sorting out some misprints there, the results fully agree.

\subsection{Next-to-leading order}
The results at next-to-leading order are a lot longer, and we decided to not
include them all explicitely. $A^L_{000}$ is however the full first order
isospin breaking amplitude (no explicit photon diagrams contribute) and
we therefore present it in App.~\ref{App:KLooo}. The expressions for the 
remaining amplitudes are available on request from the authors or can be
downloaded \cite{formulas}.

The diagrams contributing are the same as in 
\cite{BDP}, see figure there. Note however that vertices from 
${\cal L}_{W4}$ can now
also be from ${\cal L}_{W2E2}$, ${\cal L}_{S4}$ also from ${\cal L}_{S2E2}$ 
and vertices from ${\cal L}_{S2}$ or ${\cal L}_{W2}$ also from ${\cal L}_{E2}$.

The amplitudes at next-to-leading order can in principle 
include all the parameters from  ${\cal L}_{W4}$, ${\cal L}_{S4}$, 
${\cal L}_{W2E2}$ and ${\cal L}_{S2E2}$. The coefficients from the strong
and electromagnetic Lagrangians are treated as known, which leaves
$G_8$,$G_{27}$,$N^r_1,\ldots,N^r_{13}$, $D^r_1$,$D^r_2$,$D^r_4,\ldots,D^r_7$,
$D^r_{26},\ldots,D^r_{31}$ and $Z^r_1,\ldots,Z^r_{14}$ as unknown. 
In total 41 undetermined parameters. However, all of these don't have to be 
independent in the sense that they multiply the same type of term, eg. 
$m_K^4$ or $e^2 m_{\pi}^2$. It turns out that there are 30 independent 
combinations, denoted by $\tilde K_{1} \ldots \tilde K_{30}$. See 
Table~\ref{tab:isoKi} for the ones already present in the isospin limit 
\cite{BDP} 
and Table~\ref{tab:isoKi2} for the new isospin breaking combinations. 
$\tilde K_{12}$ to 
 $\tilde K_{19}$ multiply terms including $\sin\epsilon$, $\tilde K_{20}$
to $\tilde K_{28}$ terms including $e^2$ and $\tilde K_{29}$ and
$\tilde K_{30}$ terms including $Z e^2$. 

The results proportional to $G_E$ at next-to-leading order are also
published in \cite{GPS} and, after sorting out some misprints there, the
results fully agree.

The corrections to the masses and decay constants including strong isospin 
breaking were also recalculated, and they were found to agree with the  known
results, see  \cite{radiative,urech} and references therein. 
\begin{table}
\begin{center}
\begin{tabular}{|c|c|}
\hline
$\tilde K_{1}$  &  $G_8\left( N_5^r-2 N_7^r+2 N_8^r+N_9^r \right)
         +G_{27}\left(-\frac{1}{2} D_6^r \right)$\\     
$\tilde K_{2}$  &  $G_8\left(N_1^r+N_2^r  \right)
     +G_{27}\left(\frac{1}{3} D_{26}^r -\frac{4}{3}D_{28}^r\right)$\\     
$\tilde K_{3}$  &  $G_8\left(N_3^r  \right)
        +G_{27}\left(\frac{2}{3}D_{27}^r+\frac{2}{3}D_{28}^r \right)$\\     
$\tilde K_{4}$  &  $G_{27}\left(D_{4}^r-D_{5}^r+4 D_{7}^r \right)$\\     
$\tilde K_{5}$  &  $G_{27}\left(D_{30}^r+D_{31}^r+2 D_{28}^r \right)$\\
$\tilde K_{6}$  &  $G_{27}\left(8 D_{28}^r-D_{29}^r+D_{30}^r \right)$\\
$\tilde K_{7}$  &  $G_{27}\left(-4 D_{28}^r+D_{29}^r \right)$\\
$\tilde K_{8}$  &  $G_8\left(2 N_5^r+4 N_7^r+N_8^r
        -2 N_{10}^r-4 N_{11}^r-2 N_{12}^r \right)
       +G_{27}\left(-\frac{2}{3} D_1^r+\frac{2}{3} D_6^r \right)$\\     
$\tilde K_{9}$  &  $G_8\left( N_5^r+N_8^r+N_9^r \right)
                   +G_{27}\left(-\frac{1}{6} D_6^r \right)$\\     
$\tilde K_{10}$ &  $G_{27}\left(2 D_2^r-2 D_4^r-D_7^r \right)$\\ 
$\tilde K_{11}$ &  $G_{27}\left(D_7^r \right)$\\
\hline
\end{tabular}
\end{center}
\caption{The independent linear combinations of the $N_i^r$ and  $D_i^r$
that appear in the isospin invariant amplitudes.\label{tab:isoKi}}
\end{table}

\begin{table}
\begin{center}
\begin{tabular}{|c|c|}
\hline
$\tilde K_{12}$ &  $G_8 \left(3\,N_{6}^r - 4\,N_{9}^r - 5\,N_{10}^r
 - 6\,N_{11}^r - 7\,
         N_{12}^r - 6\,N_{13}^r \right)$\\ 
  & $ +G_{27} \left(23/3\,D_{1}^r - 5/3\,D_{6}^r \right)$\\
$\tilde K_{13}$ &  $G_{27}
          \left(10/3\,D_{1}^r - 11/12\,D_{6}^r \right)$\\      
$\tilde K_{14}$ &  $G_{27} \left( 5\,D_{27}^r + 5/2\,D_{31}^r \right)$\\   
$\tilde K_{15}$ &  $G_{27} \left( - 2/3\,D_{26}^r - 8/3\,D_{27}^r \right)$\\
$\tilde K_{16}$ &  $G_8 \left(2\,N_{9}^r + 3\,N_{10}^r+4\,N_{11}^r
 + N_{12}^r\right)$\\ 
 & $ +G_{27} \left(D_{1}^r - 1/3\,D_{6}^r \right)$\\
$\tilde K_{17}$ &  $G_{27} \left(D_{6}^r \right)$\\  
$\tilde K_{18}$ &  $G_8 \left(9/2\,N_{4}^r \right)
  +G_{27} \left( - 2\,D_{26}^r - 6\,D_{27}^r + D_{31}^r \right)$\\
$\tilde K_{19}$ &  $G_8 \left( 3/2\,N_{2}^r \right)
  +G_{27} \left(3/2\,D_{26}^r + 4\,D_{27}^r - D_{31}^r \right)$\\       
$\tilde K_{20}$ &  $G_8\left(- 3\,Z_{2}^r + 2\,Z_{3}^r - 18/5\,Z_{4}^r
 + 3/4\,Z_{5}^r + 
     1/2\,Z_{6}^r + 1/2\,Z_{7}^r\right.$\\  
  & $\left. + 3/4\,Z_{8}^r + 3/4\,Z_{9}^r + 1/2\,Z_{10}^r + 7/10\,Z_{11}^r
 + 7/10\,Z_{12}^r \right)$\\ 
$\tilde K_{21}$ &  $G_8 \left(2\,Z_{8}^r + 2/3\,Z_{10}^r + 2/3\,Z_{11}^r 
- 2\,Z_{12}^r \right)$\\
$\tilde K_{22}$ &  $G_8 \left( 2\,Z_{6}^r + 2\,Z_{8}^r + 2\,Z_{9}^r
 + 4/3\,Z_{10}^r +  
       2/3\,Z_{11}^r + 2/3\,Z_{12}^r + 2\,Z_{13}^r + 1/3\,Z_{14}^r\right)$\\ 
 $\tilde K_{23}$ &  $G_8 \left(Z_{12}^r \right)$\\ 
$\tilde K_{24}$ &  $G_8 \left(Z_{7}^r + 3\,Z_{8}^r
 + Z_{10}^r+3\,Z_{11}^r-Z_{12}^r \right)$\\ 
$\tilde K_{25}$ &  $G_8 \left(- 4\,Z_{1}^r - 2\,Z_{2}^r - 4/3\,Z_{3}^r
 + 4\,Z_{4}^r - 
             3/2\,Z_{5}^r+Z_{6}^r \right.$\\ 
  & $\left. -Z_{7}^r+1/2\,Z_{8}^r + 1/2\,Z_{9}^r + 1/3\,Z_{10}^r
 - 1/3\,Z_{11}^r - 1/3\,Z_{12}^r \right)$\\ 
$\tilde K_{26}$ & $G_8\left(- 12/5\,Z_{4}^r + 3/2\,Z_{5}^r - Z_{6}^r
 + Z_{7}^r-3/2\,Z_{8}^r\right.$\\ 
  & $\left.- 3/2\,Z_{9}^r - Z_{10}^r - 1/5\,Z_{11}^r 
- 1/5\,Z_{12}^r \right)$\\ 
$\tilde K_{27}$ &  $G_8 \left(8/5\,Z_{4}^r + 4/5\,Z_{11}^r 
+ 4/5\,Z_{12}^r \right)$\\ 
$\tilde K_{28}$ &  $G_8 \left(Z_{3}^r + Z_{6}^r + 3/2\,Z_{8}^r
 + 3/2\,Z_{9}^r + Z_{10}^r + 
              1/2\,Z_{11}^r + 1/2\,Z_{12}^r \right)$\\ 
$\tilde K_{29}$ &  $G_8 \left(3\,N_{10}^r + 4\,N_{11}^r + N_{12}^r \right)$\\
  & $+ G_{27} \left(D_{1}^r + 2/3\,D_{5}^r - 1/3\,D_{6}^r \right)$\\ 
$\tilde K_{30}$ &  $G_8 \left( 8/3\,N_{11}^r + 8/3\,N_{12}^r \right)
       + G_{27} \left( 16/9\,D_{5}^r + 4/9\,D_{6}^r \right)$\\ 
\hline
\end{tabular}
\end{center}
\caption{The new independent linear combinations of the $N_i^r$, $D_i^r$ and
$Z_i^r$ that appear in the isospin breaking amplitudes.\label{tab:isoKi2}}
\end{table}

\section{Numerical Results}
\label{isonumres}

\subsection{Experimental data and fit}

A full isospin limit fit was made in \cite{BDP} taking into account all
data published before May 2002. One of the reasons for this investigation 
of isospin breaking effects is to see whether isospin violation can solve 
the discrepancies in the quadratic slope parameters found there. A new full 
fit will
be done after all the electromagnetic contributions have been included
in the amplitudes (work in progress). 
The data from ISTRA+ \cite{istra} and KLOE \cite{kloe}, which 
appeared after \cite{BDP}, will then also be taken into account.

\subsection{Results with and without strong isospin breaking}

\begin{figure}
\begin{center}
\setlength{\unitlength}{1pt}
\resizebox{350 pt}{!}{
\rotatebox{270}{
\includegraphics{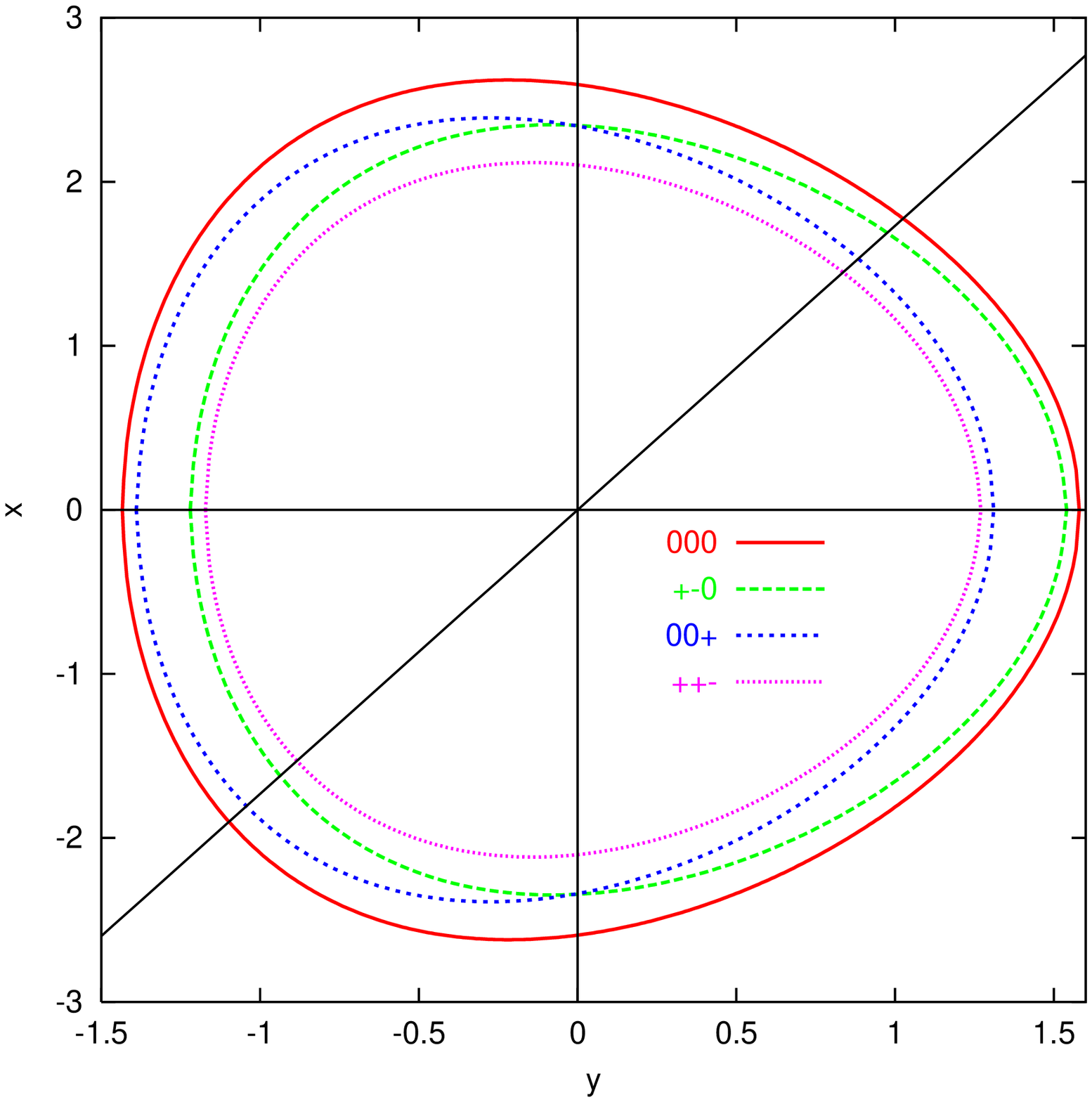}}}
\end{center}
\caption{The phase space boundaries for the five different decays 
and the curves 
along which we will compare the amplitudes.\label{fig:phase}}
\end{figure}
Our main result is the comparison between the amplitudes in the isospin limit
and including first order strong isospin breaking. 
In Fig.~\ref{fig:phase} we show the phase space boundaries for the five 
different decays and the three curves along which  we will show
results for the squared amplitudes with and without first order strong isospin 
breaking. The three curves are $x=0$, $y=0$ and $x = \sqrt{3}\,y$. 
In Fig.~\ref{fig:compare1} to Fig.~\ref{fig:compare5} we then plot the 
different squared amplitudes along these curves as a function of $r$, where
$r=\pm\sqrt{y^2+\frac{x^2}{3}}$ and the sign is chosen in the natural way, 
\be
r = \left\{ \begin{array}{ll} 
y, & x = 0 \\
x/\sqrt{3}, & y = 0 \\
y\,\sqrt{2}, & x= y\,\sqrt{3}\,.
\end{array}
\right.
\ee
Note that for all but $A^S_{+-0}$ the squared
amplitudes are normalized to their
value at the center of the Dalitz plot. A comparison of the central 
values themselves is shown in 
Table~\ref{tab:centralvalue} together with the decay rates integrated over
the Dalitz region. In Fig.~\ref{fig:compare1} one can clearly see the 
thresholds introduced by the difference between $m^2_{\pi^+}$ and
$m^2_{\pi^0}$ induced by the breaking of strong isospin invariance. These
thresholds correspond to a new process being allowed where two of 
the neutral pions are produced through an intermediate on shell state with one 
positive and one negative pion.

\begin{table}
\begin{center}
\begin{tabular}{||c|r||c|r||c|r||}
\hline
\hline
$G_8$   & $5.45$ &$L_1^r$ & $0.38\cdot10^{-3}$ & $\tilde K_1 $ & $0$ \\
$G_{27}$ & $0.392$ & $L_2^r$ & $1.59\cdot10^{-3}$ & 
$\tilde K_2/G_8 $ & $5.19\cdot 10^{-2}$ \\
$G_E$   & $-0.4$ & $L_3^r$ & $-2.91\cdot10^{-3}$ & 
$\tilde K_3/G_8 $ & $3.77\cdot 10^{-3}$ \\
$\delta_2-\delta_0$ & $-0.576$ & $L_4^r$ & $0$ & $\tilde K_4 $ & 0 \\
$\sin\epsilon$ & $1.19\cdot 10^{-2}$&$L_5^r$ & $1.46\cdot10^{-3}$ & 
$\tilde K_5/G_{27} $ & $-4.25\cdot 10^{-2}$ \\
$Z$ & $0.805$ & $L_6^r$ & $0$ & 
$\tilde K_6/G_{27} $ & $-1.66\cdot 10^{-1}$ \\
$\mu$& 0.77 GeV&$L_7^r$ & $-0.49\cdot10^{-3}$ & $\tilde K_7/G_{27} $ & 
$1.20\cdot 10^{-1}$ \\
$F_\pi$&$0.0924$ GeV& $L_8^r$ & $1.0\cdot10^{-3}$ &
  $\tilde K_8\ldots \tilde K_{11} $ & 0 \\ 
$F_K$&$0.113$ GeV&$K_1\ldots K_{11} $ & $0$ & 
$\tilde K_{12}\ldots \tilde K_{30} $ & $0$ \\
\hline
\hline
\end{tabular}
\end{center}
\caption{The various input values used.\label{tab:isoInput}}
\end{table}

The input values used to get the results can be seen 
in Table~\ref{tab:isoInput}. $L_1^r$ to $L_8^r$ come from a one-loop fit in
\cite{strongiso}, $G_8$, $G_{27}$, $\delta_2-\delta_0$ and $\tilde K_1 $ to 
$\tilde K_{11}$ from the isospin limit fit in \cite{BDP},
$G_E$ from \cite{BP} and 
$\sin\epsilon$ from \cite{strongiso}. For $Z$ we use the estimate
\be
Z = \frac{1}{2\,F_\pi^2\,e^2}\, (m_{\pi^+}^2-m_{\pi^0}^2),
\ee
which corresponds to the value in Table~\ref{tab:isoInput}. As usual, $\mu$ is 
chosen to be $0.77$ GeV. 

Very little knowledge exists of the values of $K_1\ldots K_{11}$
and $\tilde K_{12}\ldots \tilde K_{30} $, so they are set equal to
zero (tests were also made assigning order of magnitude estimates 
to them).

\begin{table}
\begin{center}
\begin{tabular}{|c|c|r||c|r|}
\cline{2-5}
\multicolumn{1}{c|}{} & 
\multicolumn{2}{|c||}{Centralvalue} & 
 \multicolumn{2}{c|}{Decay Rate} \\
\cline{2-5}
\multicolumn{1}{c|}{} & 
Isospin limit \cite{BDP} & Strong iso-br & Isospin limit \cite{BDP}& 
Strong iso-br \\
\hline
$K_L \to \pi^0 \pi^0 \pi^0$ & $6.74 \cdot 10^{-12}$ & 
$6.97 \cdot 10^{-12}$ & $2.65 \cdot 10^{-18}$ & 
$2.74 \cdot 10^{-18}$ \\
$K_L \to \pi^+ \pi^- \pi^0$ & $7.46 \cdot 10^{-13}$ & 
$7.66 \cdot 10^{-13}$ & $1.63 \cdot 10^{-18}$ & 
$1.67 \cdot 10^{-18}$ \\
$K_S \to \pi^+ \pi^- \pi^0$ & $0$ & $0$ & $3.1 \cdot 10^{-21}$ & 
$3.2 \cdot 10^{-21}$ \\
$K^+ \to \pi^0 \pi^0 \pi^+$ & $9.33 \cdot 10^{-13}$ & 
$1.01 \cdot 10^{-12}$ & $9.11 \cdot 10^{-19}$ & 
$9.84 \cdot 10^{-19}$ \\
$K^+ \to \pi^+ \pi^+ \pi^-$ & $3.72 \cdot 10^{-12}$ & 
$4.00 \cdot 10^{-12}$ & $2.97 \cdot 10^{-18}$ & 
$3.19 \cdot 10^{-18}$ \\
\hline
\end{tabular}
\end{center}
\caption{Comparison of the central values of the amplitudes squared and the
decay rates.
\label{tab:centralvalue}}
\end{table}
There are various ways to treat the masses, especially in the isospin limit 
case. In \cite{BDP} the masses used in the phase space were the physical masses
occuring in the decays. However in the amplitudes the physical kaon 
mass of the kaon involved in the process was used and the pion 
mass was given by $m_{\pi}^2=\frac{1}{3}\sum_{i=1,3} m_{\pi^i}^2$
with $i=1,2,3$ being the three pions participating in the reaction. 
This allowed for the correct kinematical relation 
$s_1+s_2+s_3= m_K^2+3m_\pi^2$ to be satisfied while having the isospin
limit in the amplitude but the physical masses in the phase space.
The results presented used the Gell-Mann-Okubo (GMO) relation for the
eta mass in the loops. Results with the physical eta mass gave small
changes within the general errors given in \cite{BDP}. 
In the  present isospin breaking calculation all the pion and kaon masses were
used correctly in the amplitude, but for the eta we again use the GMO relation,
but now including isospin breaking, 
\be
m_\eta^2 = \frac{2}{3}\, (m_{K^+}^2+m_{K^0}^2-m_{\pi^+}^2) + 
\frac{1}{3}\, m_{\pi^0}^2\,.
\ee
The possible lowest order contributions from the eta mass have been removed
from the amplitudes using the corresponding next-to-leading order relation.

Here follows a discussion of the  results in somewhat more detail. 
In general the results
are of a size as can be expected from this type of isospin breaking.
They are of order a few, up to 5\% in the amplitudes. The isospin breaking
corrections tend to increase all decay rates somewhat but this can
be compensated by small changes in the values of the fitted $\tilde K_i$
compared to the results of \cite{BDP}. The number of significant
digits quoted in Table~\ref{tab:centralvalue} is higher than the expected
precision of our results, but the trend and the general size of the change
compared to the isospin conserving results are stable with respect to
variations in dealing with the eta mass (physical or GMO).

\begin{figure}
\begin{center}
\setlength{\unitlength}{1pt}
\resizebox{350 pt}{!}{
\rotatebox{270}{
\includegraphics{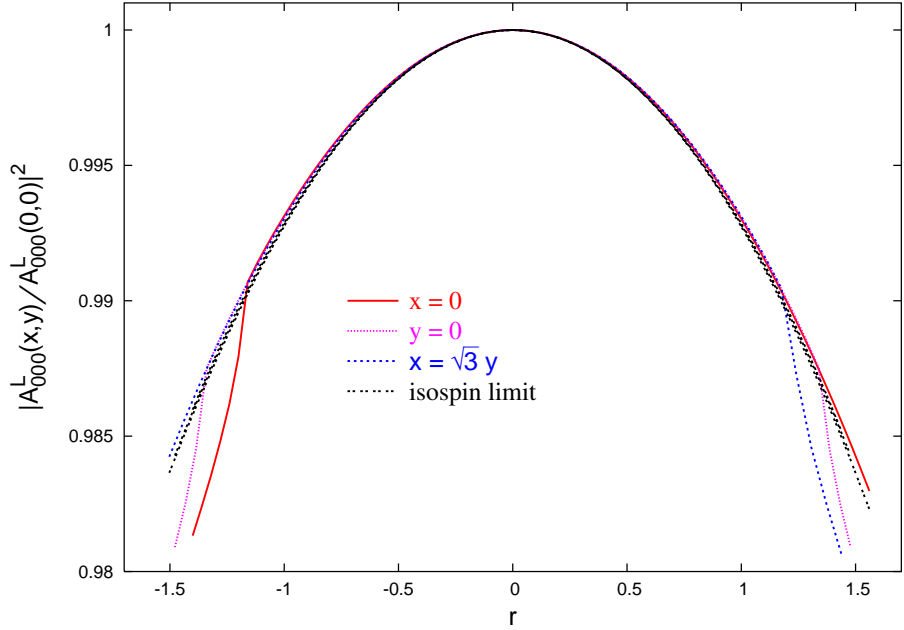}}}
\end{center}
\caption{Comparison of $K_L\to\pi^0\pi^0\pi^0$ with and without strong isospin 
breaking.}
\label{fig:compare1}
\end{figure}
For $K_L\to\pi^0\pi^0\pi^0$
the central value of the amplitude squared increases by about 3\%. The
change in the quadratic slope is similar but the total variation over
the Dalitz plot is small so the total decay rate increases by about 3\% as 
well.
This decay is the one which has most variation in the amplitude when
changing how one deals with the eta mass. The extreme case we have found was 
that this effect completely cancelled the change from isospin violation. 
For this decay the amplitude as presented in the paper
is also the final one. There are no contributions from loops with photons
for this decay. It will only be indirectly affected when the $\tilde K_i$
are determined from the decays involving charged particles which do have
contributions from loops and tree level diagrams with photons. 
Note the scale in Fig.~\ref{fig:compare1} when viewing the result.

\begin{figure}
\begin{center}
\setlength{\unitlength}{1pt}
\resizebox{350 pt}{!}{
\rotatebox{270}{
\includegraphics{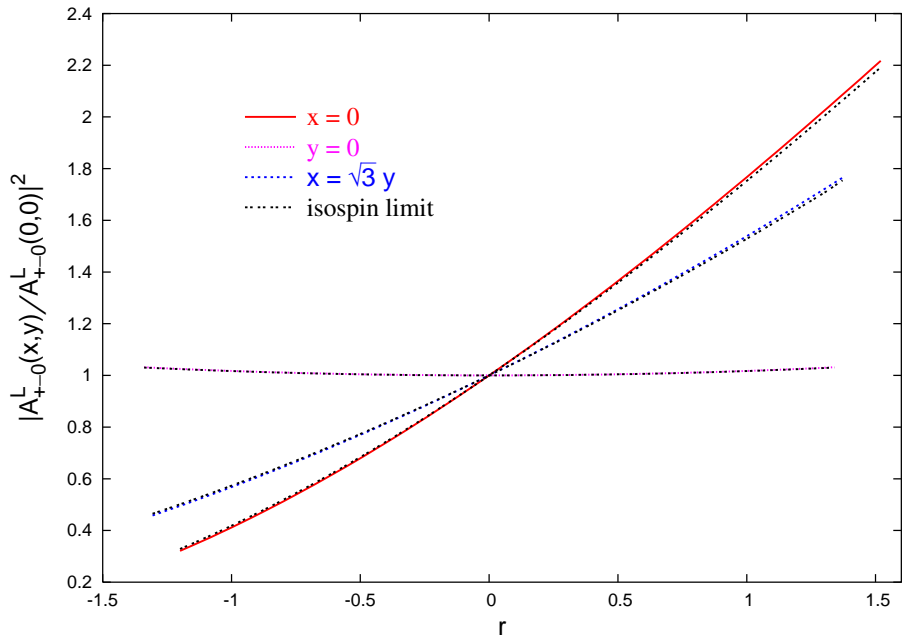}}}
\end{center}
\caption{Comparison of $K_L\to\pi^+\pi^-\pi^0$ with and without strong isospin 
breaking.}
\label{fig:compare2}
\end{figure}
The squared amplitude  $K_L\to\pi^+\pi^-\pi^0$ increases by about
2.5\% with very little
variation with the eta mass treatment. The decay rate increases by the same
amount. The changes in the Dalitz plot slopes are similar as can be
judged from Fig.~\ref{fig:compare2}.

\begin{figure}
\begin{center}
\setlength{\unitlength}{1pt}
\resizebox{350 pt}{!}{
\rotatebox{270}{
\includegraphics{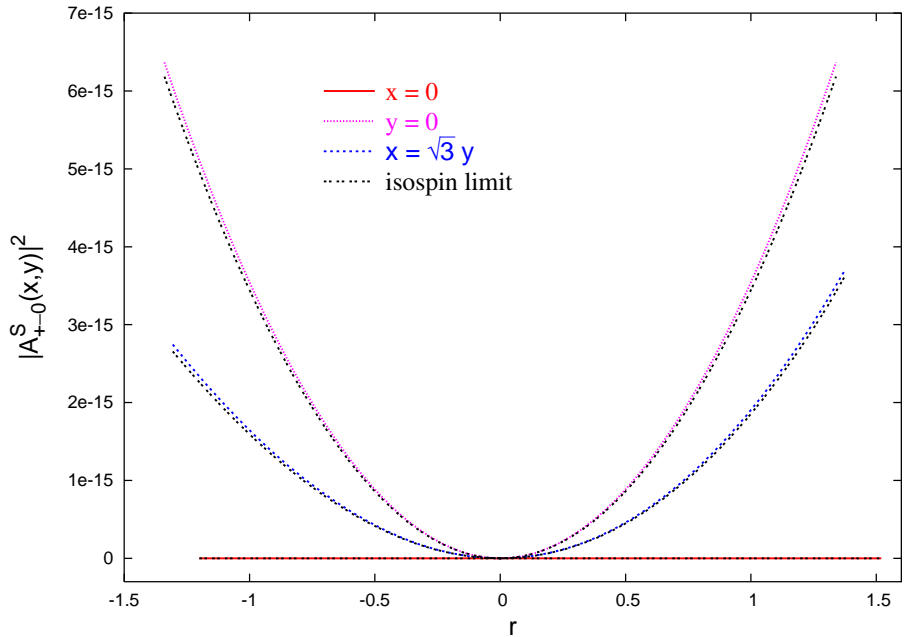}}}
\end{center}
\caption{Comparison of $K_S\to\pi^+\pi^-\pi^0$ with and without strong isospin 
breaking.}
\label{fig:compare3}
\end{figure}
For the decay $K_S\to\pi^+\pi^-\pi^0$ the amplitude in the center of the Dalitz
plot vanishes because of the symmetries. The amplitude and the slopes
increase by about 3\% as can be seen in Fig.~\ref{fig:compare3}. 

\begin{figure}
\begin{center}
\setlength{\unitlength}{1pt}
\resizebox{350 pt}{!}{
\rotatebox{270}{
\includegraphics{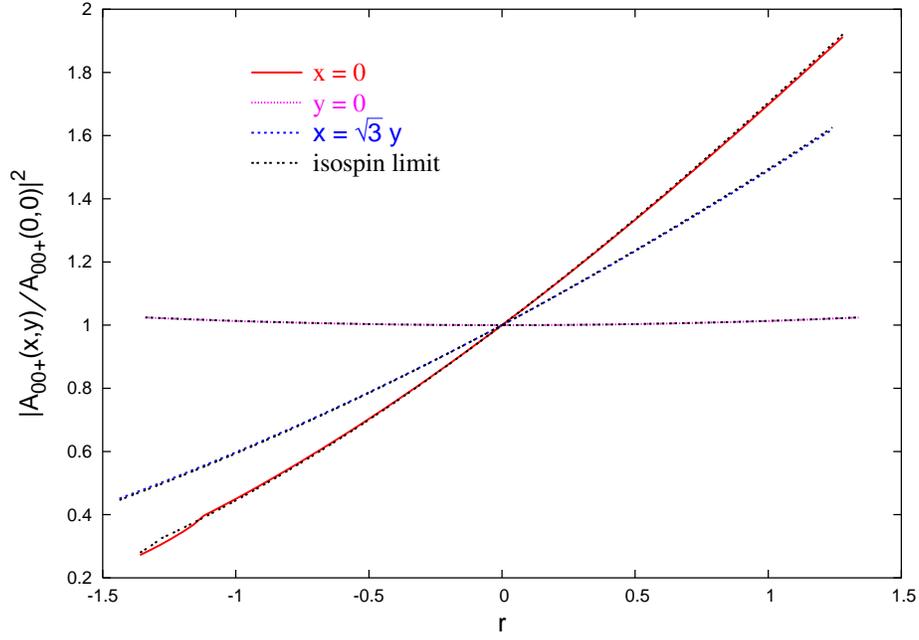}}}
\end{center}
\caption{Comparison of $K^+\to\pi^0\pi^0\pi^+$ with and without strong isospin 
breaking.}
\label{fig:compare4}
\end{figure}
The decay $K^+\to\pi^0\pi^0\pi^+$ has the largest increase.
The squared amplitude
in the center changes by about 11\%. The linear slopes decrease somewhat
leading to an increase of about 8\% to the total decay rate when compared
with the isospin conserved case.

\begin{figure}
\begin{center}
\setlength{\unitlength}{1pt}
\resizebox{350 pt}{!}{
\rotatebox{270}{
\includegraphics{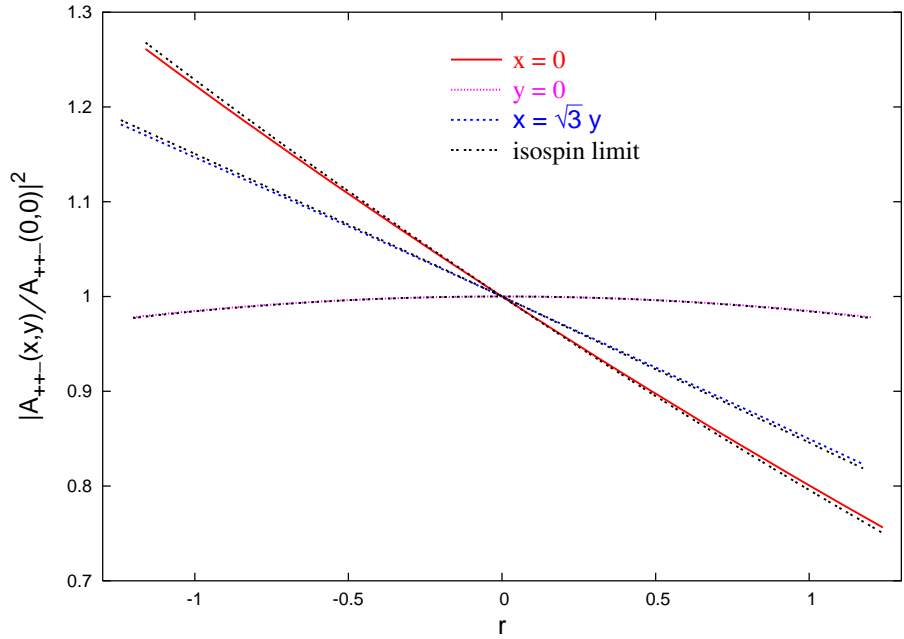}}}
\end{center}
\caption{Comparison of $K^+\to\pi^+\pi^+\pi^-$ with and without strong isospin 
breaking.}
\label{fig:compare5}
\end{figure}
The decay $K^+\to\pi^+\pi^+\pi^-$ has a change of about 7.5\% upwards in
the center of the Dalitz plot and a similar change in the decay rate.
The slopes decrease somewhat.

The conclusions above do not seem to change qualitatively when we give
the $K_i^r$ a value of about $0.001$ and the extra $\tilde K_i$ a value
relative to $G_8$ and $G_{27}$ of $0.01$. However the changes induced
by these values can be of the order of 10\%, largest for 
$K_L\to\pi^0\pi^0\pi^0$.

It should be noted that the mentioned changes are with the values of
$\tilde K_i$ determined using the isospin conserving fit. A new determination
including isospin breaking effects is planned when the diagrams with
photon propagators, both at tree level and one-loop level, have been taken
into account. At present most of the changes found can probably be
compensated by changes in the $\tilde K_i$.

\section{Conclusions}
\label{summary}
We have calculated the $K\to3\pi$ amplitudes to
next-to-leading order ($p^4,p^2\,m^2,m^4,p^2\,e^2$) in Chiral Perturbation
Theory. A similar calculation was done in \cite{BDP} in the isospin limit,
but we have now included strong isospin breaking ($m_u\neq m_d$) and local
electromagnetic isospin breaking in the amplitudes . This was done partly
because it is
interesting in general to see the possible importance of isospin breaking
in this process, but also to investigate whether isospin violation will
improve the fit to experimental data. Discrepancies between data and the
quadratic slopes from CHPT were found in \cite{BDP}, and isospin breaking may
be the cause of this.

We have tried to estimate the effects of the breaking by comparing
the squared amplitudes with and without isospin violation. The effect
seems to be at a few percent level, and probably not quite enough to solve the
dicrepancies. However, to really investigate this a new full fit has to be
done also including the explicit photon diagrams and the new data 
\cite{istra,kloe} published
after \cite{BDP}. This is work in progress
and will be presented in the future papers
Isospin Breaking in $K\to3\pi$ Decays II and III. 

\section*{Acknowledgments}
The program FORM 3.0 has been used extensively in these calculations
\cite{FORM}. This work is supported in part by the Swedish Research Council
and European Union TMR
network, Contract No. HPRN-CT-2002-00311  (EURIDICE).
\newpage

\appendix
\renewcommand{\theequation}{\Alph{section}.\arabic{equation}}
\section{The amplitude for $K_L\to \pi^0\pi^0\pi^0$.}
\setcounter{equation}{0}
\label{App:KLooo}

We divide the $M_0(s)$ function defined in Eq.~(\ref{isodefMi}) as 
\ba
M_0(s) & = & \left.M_0(s)\right|_{p^2}
+i\frac{CF_0^4}{F_\pi^3 F_K}\left\{\frac{G_8}{F_\pi^2}M_0^{8,F}(s)+ 
\frac{G_8}{F_\pi^2}\,\frac{\sin\epsilon}{\sqrt{3}}M_0^{8,\epsilon}(s) 
+ G_8\, e^2 M_0^{8,e}(s) 
\right. \nonumber \\
& + & \left. \frac{G_8^\prime}{F_\pi^2}M_0^{8\prime}(s)+
\frac{G_{27}}{F_\pi^2}M_0^{27,F}(s)+
\frac{G_{27}}{F_\pi^2}\,\frac{\sin\epsilon}{\sqrt{3}}M_0^{27,\epsilon}(s)+
G_{27}\, e^2 M_0^{27,e}(s)\right\}\,.
\ea

The effect of $G_8^\prime$ cannot be distinguished from higher order
coefficients in decays not involving external fields. This was known at tree
level earlier and has been proven to one-loop in \cite{KMW1}. This means that
terms proportinal to $G_8^\prime$ are effectively removed. 

We have extensively used the first order isospin broken GMO relation, 
\be
m_\eta^2 = \frac{2}{3}\, (m_{K^+}^2+m_{K^0}^2-m_{\pi^+}^2) + 
\frac{1}{3}\, m_{\pi^0}^2\,,
\ee
in writing the amplitude in the form below.

The explicit expressions
for  $\overline A$, $\overline B$
and $\overline B_1$, the finite part of the loop functions, can be found
in many places, e.g. \cite{ABT1}.

The octet ones are:
\ba
  \lefteqn{ M^{8,F}_0(s) =}&&
\nonumber\\&&       
\Big(2\,L^r_{1}+2\,L^r_{2}+L^r_{3}\Big)\,
\Big(  - 8\,m_{\pi^0}^4 - 8/3\,m_{K^0}^4 + 8\,s^2 \Big)
       + \Big(4\,L^r_{4}+L^r_{5}\Big)\,
\Big(  - 8/3\,m_{\pi^0}^2\,m_{K^0}^2 \Big)
\nonumber\\&&
       + \Big(N^r_{1}+ N^r_{2}+ 2\,N^r_{3}\Big)\, 
\Big( m_{\pi^0}^4 + 1/3\,m_{K^0}^4 - s^2 \Big)
       + N^r_{5}\, \Big( 10/3\,m_{\pi^0}^2\,m_{K^0}^2 + 2/3\,m_{K^0}^4 \Big)
\nonumber\\&&
       + N^r_{7}\, \Big( 4\,m_{\pi^0}^2\,m_{K^0}^2 - 4/3\,m_{K^0}^4 \Big)
       + N^r_{8}\, \Big( 10/3\,m_{\pi^0}^2\,m_{K^0}^2 + 4/3\,m_{K^0}^4 \Big)
\nonumber\\&&
       + N^r_{9}\, \Big(  - 2/3\,m_{\pi^0}^2\,m_{K^0}^2 + 2/3\,m_{K^0}^4 \Big)
       + \Big(N^r_{10}+ 2\,N^r_{11}+ N^r_{12}\Big)\,
\Big(  - 4\,m_{\pi^0}^2\,m_{K^0}^2 \Big)
\nonumber\\&&
       + \Big(\overline{A}(m_{K^0}^2)-\overline{A}(m_{K^+}^2) \Big)
\left(\frac{1/4\,m_\pi^4}{m_\pi^2-m_K^2}\right)
\nonumber\\&&
       + \overline{A}(m_{\pi^+}^2) \, \Big( 1/4\,m_{\pi^0}^2 - 1/2\,m_{K^0}^2 
\Big)
       + \overline{A}(m_{\pi^0}^2) \, \Big( 17/24\,m_{K^0}^2 \Big)
\nonumber\\&&
       + \overline{A}(m_{K^+}^2) \, \Big( 5/36\,m_{\pi^0}^2 + 25/36\,m_{K^0}^2 
\Big)
       + \overline{A}(m_{K^0}^2) \, \Big(  - 7/18\,m_{\pi^0}^2
 - 1/36\,m_{K^0}^2 \Big)
\nonumber\\&&
       +  \Big(\overline{A}(m_{K^0}^2)-\overline{A}(m_{K^+}^2) \Big)
  \, \Big( 5/36\,m_\pi^2 + 5/18\,m_K^2 \Big)
       + \overline{A}(m_\eta^2) \, \Big( 1/8\,m_{K^0}^2 \Big)
\nonumber\\&&
       + \overline{B}(m_{\pi^+}^2,m_{\pi^+}^2,s) \, \Big(  - 2\,m_{\pi^0}^2\,s 
+ m_{\pi^0}^4 + s^2 \Big)
       + \overline{B}(m_{\pi^+}^2,m_{K^+}^2,s) \,
 \Big( 5/4\,m_{\pi^0}^2\,m_{K^0}^2 
\nonumber\\&&
~~ - 5/4\,m_{\pi^0}^2\,s - 1/2\,m_{K^0}^2\,s - 1/4\,m_{K^0}^4 + 3/4\,s^2 \Big)
\nonumber\\&&
       + \overline{B}(m_{\pi^0}^2,m_{\pi^0}^2,s) \, 
\Big( 1/2\,m_{\pi^0}^2\,m_{K^0}^2 \Big)
\nonumber\\&&
       + \overline{B}(m_{\pi^0}^2,m_{K^0}^2,s) \, 
\Big( 3/8\,m_{\pi^0}^2\,m_{K^0}^2 - 1/8\,m_{K^0}^2\,s + 1/8\,m_{K^0}^4 \Big)
\nonumber\\&&
       + \overline{B}(m_{K^+}^2,m_{K^+}^2,s) \, \Big( 1/4\,m_{\pi^0}^2\,s 
+ 1/4\,m_{K^0}^2\,s - 1/4\,s^2 \Big)
\nonumber\\&&
       + \overline{B}(m_{K^0}^2,m_\eta^2,s) \, \Big( 
1/24\,m_{\pi^0}^2\,m_{K^0}^2 + 1/8\,m_{K^0}^2\,s - 5/24\,m_{K^0}^4 \Big)
\nonumber\\&&
       + \overline{B}(m_\eta^2,m_\eta^2,s) \, \Big(
 1/18\,m_{\pi^0}^2\,m_{K^0}^2 \Big)
       + \overline{B}_1(m_{\pi^+}^2,m_{K^+}^2,s) \, \Big( 
 - 1/2\,m_{\pi^0}^2\,m_{K^0}^2 + 1/2\,m_{\pi^0}^4 \Big)
\nonumber\\&&
       + \overline{B}_1(m_{\pi^0}^2,m_{K^0}^2,s) \, \Big( 
 - 1/4\,m_{\pi^0}^2\,m_{K^0}^2 + 1/4\,m_{K^0}^4 \Big)
\nonumber\\&&
       + \overline{B}_1(m_{K^0}^2,m_\eta^2,s) \, \Big(
  - 1/4\,m_{\pi^0}^2\,m_{K^0}^2 + 1/4\,m_{K^0}^4 \Big)
\nonumber\\&&
\ea
\ba
\lefteqn{ M^{8,\epsilon}_0(s) =}&&
\nonumber\\&&
\Big( 2\,L^r_{1}+ 2\,L^r_{2}+ L^r_{3}\Big)\, \Big(
  - 8\,m_\pi^4 - 8/3\,m_K^4 + 8\,s^2 \Big)
       + L^r_{4}\, \Big(  - 32/3\,m_\pi^2\,m_K^2 \Big)
\nonumber\\&&
       + L^r_{5}\, \Big(  - 40/3\,m_\pi^2\,m_K^2 + 16/3\,m_K^4 \Big)
       + L^r_{7}\, \Big( 480\,m_\pi^2\,m_K^2 - 352\,m_K^4 \Big)
\nonumber\\&&
       + L^r_{8}\, \Big( 224\,m_\pi^2\,m_K^2 - 160\,m_K^4 \Big)
\nonumber\\&&
       + \Big(N^r_{1}+N^r_{2}-2\,N^r_{3}+6\,N^r_{4}\Big)\, 
         \Big(  - m_\pi^4 - 1/3\,m_K^4 + s^2 \Big)
       + N^r_{5}\, \Big(  - 26/3\,m_\pi^2\,m_K^2 + 10\,m_K^4 \Big)
\nonumber\\&&
       + N^r_{6}\, \Big(  - 48\,m_\pi^2\,m_K^2 + 40\,m_K^4 \Big)
       + N^r_{7}\, \Big( 4\,m_\pi^2\,m_K^2 - 4/3\,m_K^4 \Big)
\nonumber\\&&
       + N^r_{8}\, \Big( 22/3\,m_\pi^2\,m_K^2 + 4/3\,m_K^4 \Big)
       + N^r_{9}\, \Big( 10/3\,m_\pi^2\,m_K^2 - 10/3\,m_K^4 \Big)
\nonumber\\&&
       + N^r_{10}\, \Big( 20\,m_\pi^2\,m_K^2 - 16\,m_K^4 \Big)
       + N^r_{11}\, \Big(  - 8\,m_\pi^2\,m_K^2 \Big)
\nonumber\\&&
       + N^r_{12}\, \Big( 68\,m_\pi^2\,m_K^2 - 64\,m_K^4 \Big)
       + N^r_{13}\, \Big( 96\,m_\pi^2\,m_K^2 - 80\,m_K^4 \Big)
\nonumber\\&&
       + \frac{1}{m_\pi^2-m_K^2}\,
 \Big(\overline{A}(m_\pi^2)-\overline{A}(m_K^2)\Big) \, \Big( m_\pi^4 \Big)
       + \overline{A}(m_\pi^2) \, \Big(  - m_\pi^2 + 13/8\,m_K^2 \Big)
\nonumber\\&&
       + \overline{A}(m_K^2) \, \Big( m_\pi^2 + 3/4\,m_K^2 \Big)
       + \overline{A}(m_\eta^2) \, \Big( 5/8\,m_K^2 \Big)
\nonumber\\&&
       + \overline{B}(m_\pi^2,m_\pi^2,s) \, \Big( 5/2\,m_\pi^2\,m_K^2
 - 12\,m_\pi^2\,s + 6\,m_\pi^4 - m_K^2\,s + 6\,s^2 \Big)
\nonumber\\&&
       + \overline{B}(m_\pi^2,m_K^2,s) \, \Big(  - 23/8\,m_\pi^2\,m_K^2
 - 7\,m_\pi^2\,s + 5\,m_\pi^4 - 3/8\,m_K^2\,s +
         27/8\,m_K^4 + 3\,s^2 \Big)
\nonumber\\&&
       + \overline{B}(m_\pi^2,m_\eta^2,s) \, \Big( 4/3\,m_\pi^2\,m_K^2
 - 4/3\,m_K^4 \Big)
\nonumber\\&&
       + \overline{B}(m_K^2,m_K^2,s) \, \Big(  - 2\,m_\pi^2\,m_K^2
 + 3\,m_\pi^2\,s + 4\,m_K^2\,s - 2\,m_K^4 - 3\,s^2\Big)
\nonumber\\&&
       + \overline{B}(m_K^2,m_\eta^2,s) \, \Big(  - 11/24\,m_\pi^2\,m_K^2
 - 3/8\,m_K^2\,s + 23/24\,m_K^4 \Big)
\nonumber\\&&
       + \overline{B}(m_\eta^2,m_\eta^2,s) \, \Big(  
- 5/18\,m_\pi^2\,m_K^2 \Big)
       + \overline{B}_1(m_\pi^2,m_K^2,s) \, \Big( 5/4\,m_\pi^2\,m_K^2
 - 5/4\,m_K^4 \Big)
\nonumber\\&&
       + \overline{B}_1(m_K^2,m_\eta^2,s) \, \Big( 3/4\,m_\pi^2\,m_K^2 
- 3/4\,m_K^4 \Big)
\nonumber\\&&
\ea
\ba
\lefteqn{ M^{8,e}_0(s) =}&&
\nonumber\\&& 
\Big(3\,Z^r_{5}- 2\,Z^r_{6}+2\,Z^r_{7}- 3\,Z^r_{8}- 3\,Z^r_{9}- 2\,Z^r_{10}
+2\,Z^r_{11}+2\,Z^r_{12}\Big)\Big( 1/9\,m_K^2 \Big)
\nonumber\\&&
       \Big(-2\, K^r_{3}+ K^r_{4}\Big)\, \Big( 1/2\,m_\pi^2 - m_K^2 \Big)
       + \Big(K^r_{5}+ K^r_{6}\Big)\, \Big( 1/3\,m_\pi^2 + 2/9\,m_K^2 \Big)
\nonumber\\&&
       + \Big(K^r_{9}+ K^r_{10}\Big)\, \Big(  - 1/3\,m_\pi^2 \Big)
      + K^r_{12}\, \Big( 8/3\,m_K^2 \Big)
\nonumber\\&&
       + \frac{m_\pi^4}{m_\pi^2-m_K^2}\,\Big(
K^r_{3}- 1/2\,K^r_{4}- 1/3\,K^r_{5}- 1/3\,K^r_{6}+ 1/3\,K^r_{9}
+ 1/3\,K^r_{10}\Big)
\nonumber\\&&
       + \frac{1}{m_\pi^2-m_K^2}\,\Big(\overline{A}(m_\pi^2)
-\overline{A}(m_K^2)\Big) \, \Big(  - m_\pi^2\,Z 
- 1/2\,m_\pi^2\,\frac{G_E}{G_8} \Big)
\nonumber\\&&
       + \frac{1}{m_\pi^2-m_K^2}\,\overline{B}(m_\pi^2,m_\pi^2,s)
 \, \Big(  - 7\,m_\pi^2\,Z\,s - 7/2\,m_\pi^2\,\frac{G_E}{G_8}\,s 
\nonumber\\&&
         ~~+ 4\,m_\pi^4\,Z + 2\,m_\pi^4\,\frac{G_E}{G_8} + 3\,Z\,s^2
 + 3/2\,\frac{G_E}{G_8}\,s^2 \Big)
\nonumber\\&&
       + \frac{1}{m_\pi^2-m_K^2}\,\overline{B}(m_\pi^2,m_K^2,s) \,
 \Big( 6\,m_\pi^2\,Z\,s + 3\,m_\pi^2\,\frac{G_E}{G_8}\,s - 4\,m_\pi^4\,Z
\nonumber\\&&
          ~~- 2\,m_\pi^4\,\frac{G_E}{G_8} - 9/4\,Z\,s^2
 - 9/8\,\frac{G_E}{G_8}\,s^2 \Big)
\nonumber\\&&
       + \frac{1}{m_\pi^2-m_K^2}\,\overline{B}(m_K^2,m_K^2,s)
 \, \Big( m_\pi^2\,Z\,s + 1/2\,m_\pi^2\,\frac{G_E}{G_8}\,s - 3/4\,Z\,s^2
          - 3/8\,\frac{G_E}{G_8}\,s^2 \Big)
\nonumber\\&&
       + \Big(\overline{A}(m_\pi^2)- \overline{A}(m_K^2)\Big)
 \, \Big(  - 1/4\,Z - 1/8\,\frac{G_E}{G_8} \Big)
\nonumber\\&&
       + \overline{B}(m_\pi^2,m_\pi^2,s) \, \Big(  - m_\pi^2\,Z
 - 1/2\,m_\pi^2\,\frac{G_E}{G_8} + Z\,s + 1/2\,\frac{G_E}{G_8}\,s \Big)
\nonumber\\&&
       + \overline{B}(m_\pi^2,m_K^2,s) \, \Big( 7/2\,m_\pi^2\,Z
 + 7/4\,m_\pi^2\,\frac{G_E}{G_8} - 3/2\,m_K^2\,Z - 3/4\,m_K^2\,\frac{G_E}{G_8}
\nonumber\\&&
    ~~      - 3/4\,Z\,s - 3/8\,\frac{G_E}{G_8}\,s \Big)
       + \overline{B}(m_K^2,m_K^2,s) \, \Big(  - 1/4\,Z\,s
 - 1/8\,\frac{G_E}{G_8}\,s \Big)
\nonumber\\&&
       + \overline{B}_1(m_\pi^2,m_K^2,s) \, \Big(  - 5/2\,m_\pi^2\,Z
 - 5/4\,m_\pi^2\,\frac{G_E}{G_8} + 1/2\,m_K^2\,Z
 + 1/4\,m_K^2\,\frac{G_E}{G_8} \Big)
\nonumber\\&&
\ea

And the 27-plets are:
\ba
\lefteqn{ M^{27,F}_0(s) =}&&
\nonumber\\&& 
\Big(2\,L^r_{1}+2\,L^r_{2}+L^r_{3}\Big)\, \Big( 8\,m_{\pi^0}^4
 + 8/3\,m_{K^0}^4 - 8\,s^2 \Big)
       + \Big(4\,L^r_{4}+ L^r_{5}\Big)\,\Big( 8/3\,m_{\pi^0}^2\,m_{K^0}^2 \Big)
\nonumber\\&&
   + \Big(D^r_{1}-3\,D^r_{2}\Big)\, \Big(  - 4/3\,m_{\pi^0}^2\,m_{K^0}^2 \Big)
       + D^r_{4}\, \Big(  - 11/3\,m_{\pi^0}^2\,m_{K^0}^2 - 1/3\,m_{K^0}^4 \Big)
\nonumber\\&&
       + D^r_{5}\, \Big(  - 1/3\,m_{\pi^0}^2\,m_{K^0}^2 + 1/3\,m_{K^0}^4 \Big)
       + D^r_{6}\, \Big( 7/9\,m_{\pi^0}^2\,m_{K^0}^2 - 1/3\,m_{K^0}^4 \Big)
\nonumber\\&&
       + D^r_{7}\, \Big(  - 10/3\,m_{\pi^0}^2\,m_{K^0}^2 - 4/3\,m_{K^0}^4 \Big)
\nonumber\\&&
       + \Big(D^r_{26}+4\,D^r_{27}-3\,D^r_{29}-6\,D^r_{30}-6\,D^r_{31}\Big)
\, \Big( 1/3\,m_{\pi^0}^4 + 1/9\,m_{K^0}^4 - 1/3\,s^2 \Big)
\nonumber\\&&
       + \frac{1}{m_\pi^2-m_K^2}\, \Big(\overline{A}(m_{K^0}^2)
-\overline{A}(m_{K^+}^2) \Big)  \, \Big( 7/12\,m_\pi^4 \Big)
\nonumber\\&&
       + \frac{1}{m_{\pi^0}^2-m_{K^0}^2}\,\overline{A}(m_{\pi^+}^2)
 \, \Big(  - 5/6\,m_{\pi^0}^4 \Big)
       + \frac{1}{m_{\pi^0}^2-m_{K^0}^2}\,\overline{A}(m_{K^+}^2)
 \, \Big( 5/3\,m_{\pi^0}^4 \Big)
\nonumber\\&&
       + \frac{1}{m_{\pi^0}^2-m_{K^0}^2}\,\overline{A}(m_{K^0}^2)
 \, \Big(  - 5/6\,m_{\pi^0}^4 \Big)
\nonumber\\&&
       + \frac{1}{m_{\pi^0}^2-m_{K^0}^2}
\,\overline{B}(m_{\pi^+}^2,m_{\pi^+}^2,s)
 \, \Big( 5/2\,m_{\pi^0}^2\,s^2-35/6\,m_{\pi^0}^4\,s + 10/3\,m_{\pi^0}^6 \Big)
\nonumber\\&&
       + \frac{1}{m_{\pi^0}^2-m_{K^0}^2}
\,\overline{B}(m_{\pi^+}^2,m_{K^+}^2,s) \, \Big(
  - 15/8\,m_{\pi^0}^2\,s^2 + 5\,m_{\pi^0}^4\,s -10/3\,m_{\pi^0}^6 \Big)
\nonumber\\&&
       + \frac{1}{m_{\pi^0}^2-m_{K^0}^2}
\,\overline{B}(m_{K^+}^2,m_{K^+}^2,s) \, \Big( 
 - 5/8\,m_{\pi^0}^2\,s^2 + 5/6\,m_{\pi^0}^4\,s \Big)
\nonumber\\&&
       + \overline{A}(m_{\pi^+}^2) \, \Big( m_{\pi^0}^2
 - 109/72\,m_{K^0}^2 \Big)
       + \overline{A}(m_{\pi^0}^2) \, \Big(  - 17/24\,m_{K^0}^2 \Big)
\nonumber\\&&
       + \overline{A}(m_{K^+}^2) \, \Big(  - 25/18\,m_{\pi^0}^2
 - 25/72\,m_{K^0}^2 \Big)
       + \overline{A}(m_{K^0}^2) \, \Big( 7/18\,m_{\pi^0}^2
 + 1/36\,m_{K^0}^2 \Big)
\nonumber\\&&
       +  \Big(\overline{A}(m_{K^0}^2)-\overline{A}(m_{K^+}^2) \Big)
  \, \Big(  - 5/36\,m_\pi^2 - 5/18\,m_K^2 \Big)
       + \overline{A}(m_\eta^2) \, \Big(  - 1/8\,m_{K^0}^2 \Big)
\nonumber\\&&
       + \overline{B}(m_{\pi^+}^2,m_{\pi^+}^2,s)
 \, \Big( 5/6\,m_{\pi^0}^2\,m_{K^0}^2 - 13/6\,m_{\pi^0}^2\,s
 + 2/3\,m_{\pi^0}^4 
\nonumber\\&&
         ~~- 5/6\,m_{K^0}^2\,s + 3/2\,s^2 \Big)
\nonumber\\&&
       + \overline{B}(m_{\pi^+}^2,m_{K^+}^2,s)
 \, \Big( 25/6\,m_{\pi^0}^2\,m_{K^0}^2 - 35/6\,m_{\pi^0}^2\,s
 + 10/3\,m_{\pi^0}^4 
\nonumber\\&&
         ~~- 53/24\,m_{K^0}^2\,s - 1/6\,m_{K^0}^4 + 19/8\,s^2 \Big)
\nonumber\\&&
       + \overline{B}(m_{\pi^0}^2,m_{\pi^0}^2,s) \, \Big(
  - 1/2\,m_{\pi^0}^2\,m_{K^0}^2 \Big)
\nonumber\\&&
       + \overline{B}(m_{\pi^0}^2,m_{K^0}^2,s) \, \Big(
  - 3/8\,m_{\pi^0}^2\,m_{K^0}^2 + 1/8\,m_{K^0}^2\,s - 1/8\,m_{K^0}^4 \Big)
\nonumber\\&&
       + \overline{B}(m_{K^+}^2,m_{K^+}^2,s) \, \Big(
  - 13/12\,m_{\pi^0}^2\,s - 7/8\,m_{K^0}^2\,s + 7/8\,s^2 \Big)
\nonumber\\&&
       + \overline{B}(m_{K^0}^2,m_\eta^2,s) \, \Big(
  - 1/24\,m_{\pi^0}^2\,m_{K^0}^2 - 1/8\,m_{K^0}^2\,s + 5/24\,m_{K^0}^4 \Big)
\nonumber\\&&
       + \overline{B}(m_\eta^2,m_\eta^2,s) \, \Big(
  - 1/18\,m_{\pi^0}^2\,m_{K^0}^2 \Big)
\nonumber\\&&
       + \overline{B}_1(m_{\pi^+}^2,m_{K^+}^2,s) \, \Big(
  - 19/12\,m_{\pi^0}^2\,m_{K^0}^2 + 1/3\,m_{\pi^0}^4 - 5/12\,m_{K^0}^4 \Big)
\nonumber\\&&
       + \Big(\overline{B}_1(m_{\pi^0}^2,m_{K^0}^2,s)
+\overline{B}_1(m_{K^0}^2,m_\eta^2,s)\Big)
 \, \Big( 1/4\,m_{\pi^0}^2\,m_{K^0}^2 - 1/4\,m_{K^0}^4 \Big)
\nonumber\\&&
\ea
\ba
 \lefteqn{ M^{27,\epsilon}_0(s) =}&&
\nonumber\\&& 
\Big(2\,L^r_{1}+2\,L^r_{2}+L^r_{3}\Big)\,\Big(
 8\,m_\pi^4 + 8/3\,m_K^4 - 8\,s^2 \Big)
\nonumber\\&&
       + L^r_{4}\, \Big( 32/3\,m_\pi^2\,m_K^2 \Big)
       + L^r_{5}\, \Big( 40/3\,m_\pi^2\,m_K^2 - 16/3\,m_K^4 \Big)
\nonumber\\&&
       + L^r_{7}\, \Big(  - 480\,m_\pi^2\,m_K^2 + 352\,m_K^4 \Big)
       + L^r_{8}\, \Big(  - 224\,m_\pi^2\,m_K^2 + 160\,m_K^4 \Big)
\nonumber\\&&
       + D^r_{1}\, \Big(  - 76/3\,m_\pi^2\,m_K^2 + 16\,m_K^4 \Big)
       + D^r_{2}\, \Big(  - 4\,m_\pi^2\,m_K^2 + 32/3\,m_K^4 \Big)
\nonumber\\&&
       + D^r_{4}\, \Big( 7/3\,m_\pi^2\,m_K^2 - 9\,m_K^4 \Big)
       + D^r_{5}\, \Big( 5/3\,m_\pi^2\,m_K^2 - 5/3\,m_K^4 \Big)
\nonumber\\&&
       + D^r_{6}\, \Big( 73/9\,m_\pi^2\,m_K^2 - 5\,m_K^4 \Big)
       + D^r_{7}\, \Big(  - 22/3\,m_\pi^2\,m_K^2 - 4/3\,m_K^4 \Big)
\nonumber\\&&
       + \Big(7\,D^r_{26}+28\,D^r_{27}-3\,D^r_{29}-6\,D^r_{30}-6\,D^r_{31}\Big)
\, \Big( 1/3\,m_\pi^4 + 1/9\,m_K^4 - 1/3\,s^2 \Big)
\nonumber\\&&
       + \frac{1}{m_\pi^2-m_K^2}\,\Big(\overline{A}(m_\pi^2)
       -\overline{A}(m_K^2)\Big) \, \Big(  - 41/6\,m_\pi^4 \Big)
\nonumber\\&&
       + \frac{1}{m_\pi^2-m_K^2}\,\overline{B}(m_\pi^2,m_\pi^2,s) \, \Big(
 25/2\,m_\pi^2\,s^2 - 175/6\,m_\pi^4\,s + 50/3\,
         m_\pi^6 \Big)
\nonumber\\&&
       + \frac{1}{m_\pi^2-m_K^2}\,\overline{B}(m_\pi^2,m_K^2,s) \, \Big(
  - 45/8\,m_\pi^2\,s^2 + 15\,m_\pi^4\,s - 10\,
         m_\pi^6 \Big)
\nonumber\\&&
       + \frac{1}{m_\pi^2-m_K^2}\,\overline{B}(m_K^2,m_K^2,s) \, \Big(
  - 55/8\,m_\pi^2\,s^2 + 85/6\,m_\pi^4\,s - 20/3
         \,m_\pi^6 \Big)
\nonumber\\&&
       + \overline{A}(m_\pi^2) \, \Big( 41/6\,m_\pi^2 + 43/12\,m_K^2 \Big)
       + \overline{A}(m_K^2) \, \Big(  - 41/6\,m_\pi^2 - 263/24\,m_K^2 \Big)
\nonumber\\&&
       + \overline{A}(m_\eta^2) \, \Big(  - 5/8\,m_K^2 \Big)
\nonumber\\&&
       + \overline{B}(m_\pi^2,m_\pi^2,s) \, \Big( 127/6\,m_\pi^2\,s
 - 38/3\,m_\pi^4 - 3/2\,m_K^2\,s - 17/2\,s^2 \Big)
\nonumber\\&&
       + \overline{B}(m_\pi^2,m_K^2,s) \, \Big(  - 101/24\,m_\pi^2\,m_K^2
 - 59/3\,m_\pi^2\,s + 40/3\,m_\pi^4+ 7/2\,m_K^2\,s
\nonumber\\&&
         ~~- 71/24\,m_K^4 + 61/8\,s^2 \Big)
       + \overline{B}(m_\pi^2,m_\eta^2,s) \, \Big(  - 4/3\,m_\pi^2\,m_K^2
 + 4/3\,m_K^4 \Big)
\nonumber\\&&
       + \overline{B}(m_K^2,m_K^2,s) \, \Big( 26/3\,m_\pi^2\,m_K^2
 - 191/12\,m_\pi^2\,s + 20/3\,m_\pi^4
\nonumber\\&&
         ~~- 311/24\,m_K^2\,s + 7\,m_K^4 + 69/8\,s^2 \Big)
\nonumber\\&&
       + \overline{B}(m_K^2,m_\eta^2,s) \, \Big( 11/24\,m_\pi^2\,m_K^2
 + 3/8\,m_K^2\,s - 23/24\,m_K^4 \Big)
\nonumber\\&&
       + \overline{B}(m_\eta^2,m_\eta^2,s) \, \Big( 5/18\,m_\pi^2\,m_K^2 \Big)
       + \overline{B}_1(m_\pi^2,m_K^2,s) \, \Big(  - 5/2\,m_\pi^2\,m_K^2
 + 25/6\,m_K^4 \Big)
\nonumber\\&&
       + \overline{B}_1(m_K^2,m_\eta^2,s) \, \Big(  - 3/4\,m_\pi^2\,m_K^2
 + 3/4\,m_K^4 \Big)
\nonumber\\&&
\ea
\ba
  \lefteqn{ M^{27,e}_0(s) =}&&
\nonumber\\&& 
 \Big(2\,K^r_{3}-K^r_{4}\Big)\, \Big(  1/2\,m_\pi^2 - m_K^2 \Big)
       +\Big(K^r_{5}+ K^r_{6}\Big)\, \Big(  - 1/3\,m_\pi^2 - 2/9\,m_K^2 \Big)
\nonumber\\&&
       + \Big(K^r_{9}+ K^r_{10}\Big)\, \Big( 1/3\,m_\pi^2 \Big)
       + K^r_{12}\, \Big(  - 8/3\,m_K^2 \Big)
\nonumber\\&&
       + \frac{m_\pi^4}{m_\pi^2-m_K^2}\,\Big(-K^r_{3}
 +1/2\,K^r_{4}+1/3\,K^r_{5}+1/3\,K^r_{6}-1/3\,K^r_{9}-1/3\,K^r_{10}\Big)
\nonumber\\&&
       + \frac{1}{m_\pi^2-m_K^2}\,\Big(
     \overline{A}(m_\pi^2)-\overline{A}(m_K^2)\Big) 
      \, \Big(  - 2/3\,m_\pi^2\,Z \Big)
\nonumber\\&&
       + \frac{1}{m_\pi^2-m_K^2}\,\overline{B}(m_\pi^2,m_\pi^2,s)
 \, \Big(  - 14/3\,m_\pi^2\,Z\,s + 8/3\,m_\pi^4\,Z + 2\,Z\,s^2 \Big)
\nonumber\\&&
       + \frac{1}{m_\pi^2-m_K^2}\,\overline{B}(m_\pi^2,m_K^2,s)
 \, \Big( 4\,m_\pi^2\,Z\,s - 8/3\,m_\pi^4\,Z - 3/2\,Z\,s^2 \Big)
\nonumber\\&&
       + \frac{1}{m_\pi^2-m_K^2}\,\overline{B}(m_K^2,m_K^2,s)
 \, \Big( 2/3\,m_\pi^2\,Z\,s - 1/2\,Z\,s^2 \Big)
\nonumber\\&&
       + \Big( \overline{A}(m_\pi^2)-\overline{A}(m_K^2)\Big)  
        \, \Big(  - 1/6\,Z \Big)
       + \overline{B}(m_\pi^2,m_\pi^2,s)
 \, \Big(  - 2/3\,m_\pi^2\,Z + 2/3\,Z\,s \Big)
\nonumber\\&&
       + \overline{B}(m_\pi^2,m_K^2,s)
 \, \Big( 7/3\,m_\pi^2\,Z - m_K^2\,Z - 1/2\,Z\,s \Big)
       + \overline{B}(m_K^2,m_K^2,s) \, \Big(  - 1/6\,Z\,s \Big)
\nonumber\\&&
       + \overline{B}_1(m_\pi^2,m_K^2,s) \, \Big(
  - 5/3\,m_\pi^2\,Z + 1/3\,m_K^2\,Z \Big)
\nonumber\\&&
       - 14/3\,m_K^2\,Z\left(\frac{1}{\epsilon}+\ln(4\pi)+\Gamma^\prime+1
\right) \,\frac{1}{16\pi^2}\,.
\nonumber\\&&
 \ea
The last term is the infinity remaining from the loops since we haven't 
included
the ${\cal O}(G_{27}\, e^2 p^2)$ Lagrangian.

\end{document}